\documentclass[aps,prb,reprint,superscriptaddress]{revtex4-1}

\usepackage[english]{babel}
\usepackage{graphicx}
\usepackage{ifpdf}
\usepackage{bm}	
\usepackage{latexsym}
\usepackage{color,soul}
\usepackage{pstricks}
\usepackage{figsize}
\usepackage{float}
\usepackage{color}
\usepackage{amssymb}
\usepackage{hyperref}
\usepackage{amsmath}
\usepackage{epstopdf}
\usepackage{verbatim}
\usepackage{bbold}

\begin{document}

\title{Effect of interactions and disorder on the relaxation of two-level systems in amorphous solids}

\author{Ofek Asban}
\affiliation{Department of Physics, Ben-Gurion University of the
Negev, Be’er-Sheva 84105, Israel}

\author{Ariel Amir}
\affiliation{School of Engineering and Applied Sciences, Harvard University, Cambridge, Massachusetts, 02138, USA}

\author{Yoseph Imry}
\affiliation{Department of Condensed Matter Physics, Weizmann Institute of Science, Rehovot 76100, Israel}

\author{Moshe Schechter}
\affiliation{Department of Physics, Ben-Gurion University of the
Negev, Be’er-Sheva 84105, Israel}

\date{\today}

\begin{abstract}
At low temperatures the dynamical degrees of freedom in amorphous solids are tunneling two-level systems (TLSs). Concentrating on these degrees of freedom, and taking into account disorder and TLS-TLS interactions, we obtain a "TLS-glass," described by the random field Ising model with random $1/r^3$ interactions. In this paper we perform a self consistent mean field calculation, previously used to study the electron-glass (EG) model [A.~Amir {\it et al.}, Phys. Rev. B {\bf 77}, 165207, (2008)]. Similarly to the electron-glass, we find a $\frac{1}{\lambda}$ distribution of relaxation rates $\lambda$, leading to logarithmic slow relaxation. However, with increased interactions the EG model shows slower dynamics whereas the TLS glass model shows faster dynamics. This suggests that given system specific properties, glass dynamics can be slowed down or sped up by the interactions.
\end{abstract}


\maketitle
\section{Introduction}

At low temperatures amorphous solids show anomalous behavior with respect to their ordered counterparts.
As was first noted by Zeller and Pohl \cite{Zeller&Pohl1971} the equilibrium properties of amorphous solids have different temperature dependence than predicted by the Debye model; some examples are the temperature dependences of the heat capacity
$c_v \propto T^{\alpha}$, and the thermal conductivity $\kappa \propto T^{\beta}$ where $\alpha \approx 1$ and $\beta \approx 2$. Moreover, phonon attenuation is qualitatively and quantitatively universal in a large variety of disordered and amorphous materials. Shortly after, Anderson {\it et al.} \cite{Anderson&Halperin&Varma} and Phillips \cite{PhillipsDistributionOfTLSs} independently developed the standard tunneling model (STM), a phenomenological model which quite successfully accounts for many of the low temperature characteristics of amorphous solids. The STM states that at low temperatures the dominant dynamical degrees of freedom are two-level systems (TLSs); each TLS represents an atom or a group of atoms that occupy one of two localized configuration states that result from an asymmetric double-well potential. TLSs are defined by their asymmetry energy $\Delta$ and tunneling amplitude $\Delta_0 \sim e^{-\Lambda}$. Given the random nature of the system, $\Delta$ and $\Lambda$ are assumed to be distributed uniformly leading to the distribution $P(\Delta, \Delta_0)=\frac{P_0}{\Delta_0}$\cite{Anderson&Halperin&Varma, PhillipsDistributionOfTLSs,PhillipAmorphousLowTemp}. TLSs reach thermal equilibrium with the phonon bath through a linear coupling to the local strain. Whereas in its basic form the STM considers noninteracting TLSs, TLSs interact via acoustic and electric dipole interactions. TLS-TLS interactions result in, e.g., spectral diffusion \cite{Black&HalperinTLS-TLSInt}, delocalization of low energy pair excitations \cite{burinCrossOverTLS-TLSRelaxation}, and slow relaxation of dielectric and acoustic response at very low temperatures \cite{Rogge&Natelson&Osheroff, Natelson&Rosenberg&Osheroff} suggesting the formation of a TLS-glass (TG).

Recent work on microfabricated devices caused a renewed interest in TLSs, both for harnessing them for technological applications, for example quantum memory \cite{Neeley2008}, and for avoiding their destructive influence as a source of noise. In particular, superconducting quantum bits (qubits) have shown extreme sensitivity to even a single TLS \cite{Simmonds2004,Barends2014}. This coupling of the qubit system to TLSs was then used to investigate the characteristics of individual TLSs \cite{Neeley2008, Shalibo2010, lisenfeldDecoherenceTLSs, Matityahu&Shnirman&Schechter} and specifically the nature of TLS-TLS interactions up to the accuracy of a single interacting pair \cite{Lisenfeld2015}.

The thermodynamic and the dynamic properties of single non-interacting TLSs have been studied thoroughly \cite{Anderson&Halperin&Varma,PhillipAmorphousLowTemp,LisenfeldTDependenceOfSingleTLS,Grabovskij&LisenfeldStrainTuningOfSingleTLS,Muller&LisenfeldNoiseFromSingleTLS,Matityahu&Shnirman&Schechter}.
The many-body dynamics of interacting TLSs is, however, more complicated. It was studied e.g. with regard to the relaxation of the dielectric response of the interacting TLSs system \cite{Burin'sDipoleGap,Rogge&Natelson&Osheroff,Natelson&Rosenberg&Osheroff,Burin1998} and to the relaxation and decoherence of resonant TLS pairs \cite{burinCrossOverTLS-TLSRelaxation,Burin&Kagan}. In this paper we are interested in the relaxation dynamics of the occupations of the interacting TLS-glass in a large parameter regimes, and in its dependence on the strength of the interaction, strength of disorder, and system size.

To obtain the relaxation dynamics of the TLS glass we follow a similar method to that used previously for the electron-glass (EG) model\cite{Ariel'sPaper}. We calculate numerically the density of states of the interacting TLS system in the mean-field approximation and use it to determine the TLS-phonon transition rates. The total relaxation of the system is then calculated by taking the norm of the occupation vector, which is the solution of the linearized Pauli rate equation. Taking the rates to the continuum and using the $1/\lambda$ distribution of rates the logarithmic slow relaxation is obtained. Furthermore, the logarithm depends on interactions and disorder through the minimum cutoff rate, $\lambda_{min}$. Using this dependence we examine the qualitative effect of the interactions, disorder and systems size on the dynamics, and compare it with the EG model.

The structure of the paper is as follows: In Sec.~\ref{The TLS-Glass model} we define the local-equilibrium state of the system, present the model in the mean-field approximation and obtain numerically the single particle density of states (DOS) which contains the dipole gap. In Sec.~\ref{Dynamics}, we derive the logarithmic shape of the relaxation. In Sec.~\ref{Results} we show the numerical results of the DOS and the distribution of rates for different values of the disorder and interaction. In Sec.~\ref{Comparison to the EG model} we compare our results to the results of the EG model under the same schemes of parameter variation, and discuss the dependence of the relaxation on the system size for both the EG and TLS models. We then conclude in Sec.~\ref{conclusions}.

\section{The TLS-Glass model}
\label{The TLS-Glass model}
In this section we discuss the STM model the with addition of Ising type interactions between the TLSs. We then apply the mean-field (MF) approximation and obtain the self-consistent equations.

We consider the Hamiltonian
\begin{equation}
\begin{split}
\label{eq: The Full Model}
\mathcal{H}_{TG} = &\sum_i \left(\Delta_i S^z_i + \Delta_{0i} S^x_i\right) - \frac{1}{2}\sum_{i \neq j} \frac{u_{ij}}{r^3_{ij}} S^z_i S^z_j \, , \\
\end{split}
\end{equation}
where $S^{z(x)}=\frac{1}{2}\sigma^{z(x)}$ represent the TLSs ($\sigma^{z(x)}$ are the Pauli matrices).
$J_{ij}=\frac{u_{ij}}{r_{ij}^3}$ represents both acoustic and electric interactions between TLSs. Since both interactions depend on the orientations and relative positions of the two TLSs, we choose $u_{ij}$ from a random Gaussian distribution.
\begin{equation}
\label{u distribution}
p(u)=\frac{1}{\sqrt{2 \pi}U_0} \exp\left(-\frac{1}{2}\frac{u^2}{U_0^2}\right) \, ,
\end{equation}
and quantify the interaction strength by $J \equiv U_0/r^3_{nn}$, where $r_{nn}$ is the average nearest-neighbor distance. Numerically we set $r_{nn}=1$.

To obtain the MF energies one can apply on $\mathcal{H}_{TG}$, Eq.~(\ref{eq: The Full Model}), a variational derivative with respect to $S^z_i$ and obtain the MF asymmetry energy $\Delta'_i$ ,
\begin{equation}
\label{eq: MF TLS}
\Delta'_i = \frac{\delta \mathcal{H}_{TG}}{\delta S^z_i} = \Delta_i - \frac{1}{2}\sum^N_{j(\neq i)}\frac{u_{ij}}{r^3_{ij}}S^z_j
\end{equation}
where $\Delta_i$ is the asymmetry energy of TLS $i$. $\Delta_i$ is chosen from a Gaussian distribution with variance $W$, which we use to quantify the disorder. $N$ is the number of sites (system size).

After thermal averaging the obtained self-consistent equation (SCE) is:
\begin{equation}
\label{eq: MF TLS equilibrium}
\Delta'_i = \Delta_i + \frac{1}{4}\sum_{j\neq i}\frac{u_{ij}}{r^3_{ij}}\tanh\left(\frac{\Delta'_j}{2T}\right)
\end{equation}
where we set the Boltzmann constant to unity, reassign $\Delta'_i = \langle \Delta'_i \rangle_T$, and use $\langle S^z_i \rangle_T = \frac{1}{2}\langle \sigma^z_i \rangle_T = -\frac{1}{2}\tanh\left(\frac{1}{2}\beta \Delta'_i \right)$.
The single TLS shifted Hamiltonian is:
\begin{equation}
\label{eq: Single TLS Hamiltoinian}
\mathcal{H}'_{TLS} = \sum_i \left(\Delta'_i S^z_i + \Delta_{0i} S^x_i\right)
\end{equation}
and the equilibrium excitation energy of the $i$'th TLS is:
\begin{equation}
\label{eq: MF TLS diagonal}
E_i =  \text{sgn}(\Delta'_i)\sqrt{{\Delta'_i}^2+\Delta^2_{0i}}
\end{equation}

Unlike the distribution given in the STM, $p(\Delta, \Delta_0)=\frac{P_0}{\Delta_0}$, which is uniform in the asymmetry energies, we choose
\begin{equation}
\label{TLS distribution 2}
p\left(\Delta, \Delta_0\right) = \frac{P_0}{\Delta_0} \frac{1}{\sqrt{2\pi}W^2}\exp\left(-\frac{1}{2} \frac{\Delta^2}{W^2}\right) \, .
\end{equation}
This choice eventually does not affect the qualitative physical outcome. However, it allows us to look at the effect of changing disorder. It is also in line with the DOS of the asymmetry energies for the relevant TLSs at low energies in KBR:CN (CN flips)\cite{ChurkinTLSDOSMonteCarlo}, as well as in the two TLS model\cite{MosheStrong&WeakTLSs,churkin2013strain}.

\begin{figure}[ht]
	\centering
	\includegraphics[scale=0.7]{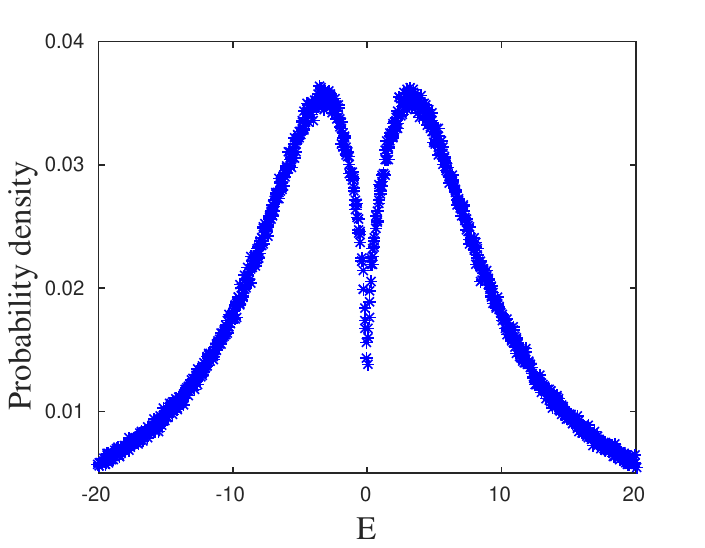}
	\caption{(Color online) TLS DOS. The normalized histogram (DOS) of TLS energies $E_i = \text{sgn}(\Delta')\sqrt{{\Delta'}_i^2+\Delta_{0i}^2}$ obtained by solving the self-consistent equations, Eq.~(\ref{eq: MF TLS}), for $N=10000$ sites, $W=J=1, T=0.05$. Results are averaged over 300 realizations.}
\label{fig: tls dos}
\end{figure}

Previous work \cite{CoulombAndDipoleGapsDerivation,CuevasTLSDOSMonteCarlo,Burin'sDipoleGap} has shown that the DOS of the TG system in 3D has a logarithmic gap which results from the dipole interactions. We present the numerical solution of the self-consistent equations (Eq.~(\ref{eq: MF TLS})), which gives the same logarithmic dependence, and in addition the behavior for larger energy values far from the gap region. Note, however, that for large disorder the gap width is exponentially small in the parameter $W/J$\cite{MosheStrong&WeakTLSs}, unlike the polynomial dependence on disorder for the EG model \cite{EfrosShklovskiiCoulombGapT=0}. The calculation of the TG energies within mean field allows us to gain an understanding of the relation between the DOS and the dynamics of the system, and in particular its dependence on control parameters of the model such as disorder, interaction strength and system size.
Following an iterative procedure done by Grunewald {\it et al.} \cite{NumericalMethodForDOS} we calculate numerically the solution of the SCE, Eq.~(\ref{eq: MF TLS equilibrium}), for finite temperature.
We set the initial values of the MF asymmetry energy $\Delta'_i$ to uniform distribution, and perform an iterative procedure that eventually converges to  the solution of the SCE. We then obtain the excitation energies of the TLSs given in Eq.~(\ref{eq: MF TLS diagonal}). The normalized histogram (DOS) of the energies is plotted in Fig.~\ref{fig: tls dos}.

\begin{figure}[ht]
	\centering
	\includegraphics[scale=0.7]{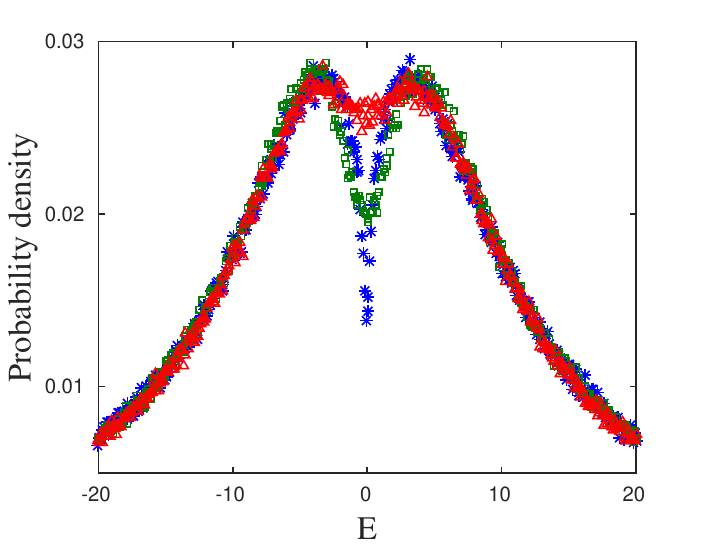}
	\caption{(Color online) TLS DOS for different temperatures. $T=0.1$ (blue asterisks), $T=1$ (green squares), and $T=2$ (red triangles) for $N=1000$. The gap gradually disappears with increasing temperature.}
\label{TLSDOSDiffT}
\end{figure}
Furthermore, as shown numerically in Fig.~\ref{TLSDOSDiffT} the gap disappears gradually as the temperature increases. A similar phenomenon occurs for the electron-glass model (as discussed in Sec.~\ref{Comparison to the EG model}). Note that $\Delta_0$'s do not evolve with the iterations since their coupling to phonons is neglected \cite{Anderson&Halperin&Varma, PhillipsDistributionOfTLSs}.
In all the numerical calculations the parameters of Eq.~(\ref{eq: MF TLS}) are measured in units of the interaction at average nearest-neighbor distance $J=\frac{U_0}{r^3_{nn}}$ and the TLSs are distributed homogeneously in a three dimensional cube with periodic boundary conditions. Also, $\Delta_0$ is taken to be in the range $[10^{-7}, 10^{-1}]$ \cite{Burin'sDipoleGap, Anderson&Halperin&Varma}. Excluding the case where the interaction parameter, $J$, is varied explicitly, we set the tunneling strength to be $\chi \equiv P_0U_0 = 10^{-3}$ given the fact that it ranges between $10^{-3}$ and $10^{-4}$ in all known amorphous materials \cite{PhillipAmorphousLowTemp}.

\section{Dynamics}
\label{Dynamics}

In this section we follow a similar approach to that used by
Amir {\it et al.} \cite{Ariel'sPaper} for the EG, and obtain the relaxation to local equilibrium of the TG model. The dynamics of the average occupation of state $i$ at time $t$ is generally described by the
Pauli master equation:
\begin{equation}
\label{eq: Master-Equation general}
\frac{d p_i(t)}{d t} = \sum_{j\neq i} \omega_{ij} p_j(t) - \omega_{ji} p_i(t)
\end{equation}
where $\omega_{ij}$ is the transition rate from state $j$ to state $i$ and the occupation $p_i(t)$ can take the values in the range $[0,1]$.
Equation (\ref{eq: Master-Equation general}) conserves the total probability; i.e., $\sum_{i=1} p_i(t)$ is constant. Specifically for the EG, \cite{Ariel'sPaper} this reflects the conservation of the total number of electrons.
However, in the TG system the transition of probability between any two TLSs is not allowed and therefore there is probability conservation for each TLS separately, $\sum_{m=1,2} p^i_{m}(t) = p^i_{1} + p^i_{2} = 1$, where $p^i_1, p^i_2$ are respectively the average probability occupations of the low energy and high energy local levels of the TLS at site $i$. Accordingly, Eq.~(\ref{eq: Master-Equation general})
is reduced to two coupled rate equations of the occupations of the $i$'th TLS:
\begin{equation}
\begin{split}
\label{eq: Master-Equation TLS with p}
\frac{d p^i_1(t)}{d t} &= \omega^i_- p^i_2(t) - \omega^i_+ p^i_1(t)\\
\frac{d p^i_2(t)}{d t} &= \omega^i_+ p^i_1(t) - \omega^i_- p^i_2(t)
\end{split}
\end{equation}
where $\omega^i_{+}$ and $\omega^i_{-}$ are respectively the TLS upward and downward transition rates caused by the TLS interaction with the phonon bath
$\sum_i\sum_k g_{ik}\left(a^{\dagger}_{-k} + a_{k}\right) S^x_i$, where $k$ represents phonon with momentum vector $\bm{q}$ and polarization $s$, and $g_{ik}$ is coupling constant which is proportional to the deformation potential constant $\gamma_{is}$. The rates are obtained via Fermi's golden rule
\cite{PhillipAmorphousLowTemp}:
\begin{equation}
\label{DownwardTransition}
\omega^i_{-} = \sum_s\frac{\gamma_{is}^2}{c^5_s} \frac{\Delta_{0i}^2 E_i}{2\pi\rho\hbar^4}\left(N_i+1\right) \equiv a_i \Delta_{0i}^2 E_i \left(N_i+1\right)
\end{equation}
and a similar expressions for $\omega^i_{+}$, with the brackets in Eq.~(\ref{DownwardTransition}) replaced with $N_i$. Here $a_i \equiv \sum_s\frac{\gamma_{is}^2}{c^5_s 2\pi\rho\hbar^4} \simeq 10^8K^{-3}s^{-1}$, where Boltzmann constant is set to unity, $N_i=\left(e^{\beta E_i}-1\right)^{-1}$ is the equilibrium phonon occupation at a given energy splitting of the TLS $(E_i)$, and $\beta = \frac{1}{T}$ is the inverse temperature.
Finally, Eq.~(\ref{eq: Master-Equation TLS with p}) reduces to one parameter in the pseudospin representation by substituting $\sigma_i = \langle\sigma^z_i\rangle = p^i_{2} - p^i_{1}$:
\begin{equation}
\label{eq: Master-Equation TLS sigma rep}
\frac{d\sigma_i}{dt} = -2a_i \Delta_{0i}^2 E_i \left[\sigma_i\left(N_i+\frac{1}{2}\right)+\frac{1}{2}\right] = -\lambda_i \sigma_i - a_i\Delta_{0i}^2 E_i
\end{equation}
where the TLS-phonon relaxation rate in equilibrium is \cite{JakleTransitionRatesOfTLSs, PhillipAmorphousLowTemp}
\begin{equation}
\label{relaxation rate}
\lambda_i = -(\omega^i_- + \omega^i_+) = -a_i\Delta^2_{0i} E_i\coth\left(\frac{E_i}{2T}\right) \, .
\end{equation}
Equation (\ref{eq: Master-Equation TLS sigma rep}) has a simple form but has hidden complexity. The right-hand side depends on the energy $E_i$ which in turn depends on the interactions, disorder and out-of-equilibrium occupations of all the TLSs in the system, i.e., $E_i(\bm{\sigma'})$ where $\bm{\sigma'}$ denotes all the elements of the pseudospin vector except the $i$th element.

For TLS occupations slightly out of equilibrium ($\delta \sigma_i \equiv \sigma_i - \sigma_i^0 \ll 1$) we can expand the right-hand side of Eq.~(\ref{eq: Master-Equation TLS sigma rep}) to first order in $\delta\sigma_i$ around the local equilibrium point. Neglecting a subdominant interaction term\cite{Ariel'sPaper} (see App.~\ref{Appendix_2} for details) we obtain
\begin{equation}
\label{eq: rate equation TLS up to 1st order}
\frac{d\delta\sigma_i}{dt} \simeq - \lambda_i\delta\sigma_i \, .
\end{equation}
Eq.~\ref{eq: rate equation TLS up to 1st order} has a simple solution:
\begin{equation}
\label{eq: Master-Equation solution}
\delta\sigma_i(t) = c_i e^{-\lambda_i t}
\end{equation}
where $c_i\equiv \delta \sigma_i(0)$ is the initial deviation of TLS $i$ at the moment the external strain driving force has stopped. To quantify the total relaxation of the system one can take the norm of the vector $\bm{\delta\sigma}$ \cite{Ariel'sPaper}:
\begin{equation}
\label{eq: Total relaxation TLS discrete}
|\bm{\delta \sigma}| = \sum_i c_i e^{-\lambda_i t}
\end{equation}
and in the continuous limit,
\begin{equation}
\label{eq: Total relaxation TLS}
|\bm{\delta\sigma}| = c\int^{\lambda_{max}}_{\lambda_{min}} p(\lambda) e^{-\lambda t} d\lambda \,
\end{equation}
where a uniform distribution of initial excitations $c(\lambda) = c$ is assumed\cite{Ariel'sPaper}. The rate distribution is then calculated numerically and obeys a $\frac{1}{|\lambda|}$ distribution over a very broad rate regime.

\begin{figure}[ht]
\centering
\begin{subfigure}
  \centering
  \includegraphics[scale=0.7]{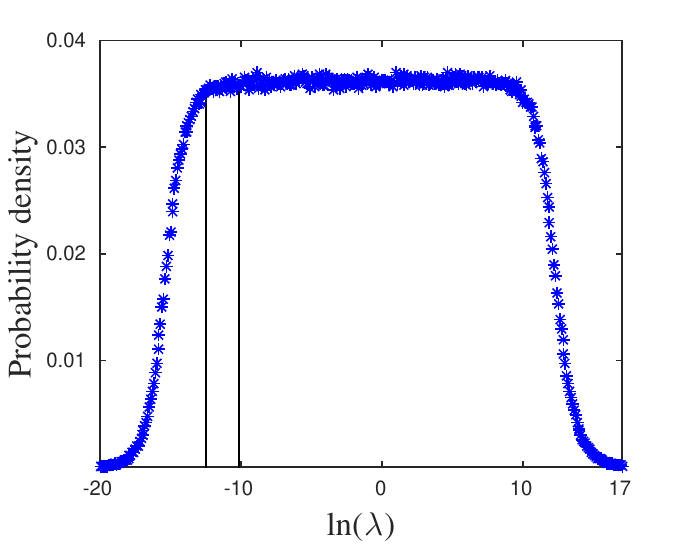}
\end{subfigure}
\begin{subfigure}
  \centering
  \includegraphics[scale=0.7]{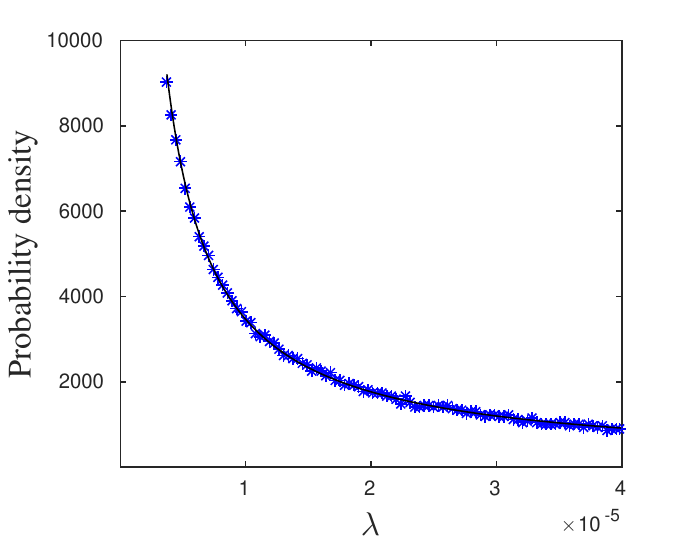}
\end{subfigure}
\caption{(Color online) Decay rate distribution for the TLS-glass. The distribution of decay rates $\lambda_i$ presented in Eq.~(\ref{relaxation rate}) calculated for $N=10000$ and $\frac{J}{T}=10$. The disorder energy, tunneling splinting and interaction strength are the same as in Fig.~\ref{fig: tls dos}. The graph is averaged over 1000 realizations. (a) Rate distribution in log scale for the full range of values. The bulk of rates occupy the plateau region which translates to $\frac{1}{\lambda}$ form in normal scale. (b) Rate distribution within the cutoff region in the log plot shown in linear scale with a $\frac{1}{\lambda}$ fit.}
\label{fig: TLS Rate}
\end{figure}

In Fig.~\ref{fig: TLS Rate} we plot the rates distribution using Eq.~(\ref{relaxation rate}) and the energies given in Fig.~\ref{fig: tls dos}. The normal scale is shown in a regime determined by a lower cutoff being the minimum value of the plateau region in the log plot, $\lambda_{min}$. This value also determines the relaxation time scale of the system (see Eq.~(\ref{eq: log relaxation TLS}) below). The maximum cutoff value $\lambda_{max}$ is fixed arbitrarily and has no significance.
The $\frac{1}{|\lambda|}$ functional form of the distribution is a result of the dependence of the decay rates on $\Delta^2_0$ [see Eq.~(\ref{relaxation rate})] in conjunction with the joint distribution function $p(\Delta_0, \Delta') \propto 1/\Delta_0$.
Finally we substitute in Eq.~(\ref{eq: Total relaxation TLS}) the rate distribution and obtain the logarithmic relaxation:
\begin{equation}
\begin{split}
\label{eq: log relaxation TLS}
|\bm{\delta\sigma}| \simeq -c\left[\gamma_E+log\left( \lambda_{min} t\right)\right]
\end{split}
\end{equation}
where $\gamma_E\approx 0.577$ is the Euler constant and the integral is approximated for $1/\lambda_{max} < t < 1/\lambda_{min}$ \cite{ExpInt}. The logarithmic relaxation we find here is in line with the TLS-glass being a part of a large class of materials with a similar slow logarithmic relaxation \cite{AmirPnas2012}.
As mentioned above, the $\frac{1}{|\lambda|}$ functional form of the rates distribution is dominated by the distribution of $\Delta_0$, and is thus independent of disorder and interaction strengths. However, the latter change the density of states of single particle excitations, and specifically that at low energies, that dominate the slowest transition rates \cite{PhillipAmorphousLowTemp}. Thus, disorder and interaction can shift the distribution of relaxation rates to higher or lower values, see below.

\section{Effect of interactions and disorder}
\label{Results}

In this section we study the effect of the variation of disorder and interaction on the DOS and on the dynamics of the TLS glass. In particular, we find that the interactions speed up the relaxation process rather than slow it down, in contrast to what was found for the EG model \cite{Ariel'sPaper} (see also Fig.~\ref{fig: EG Rate diffInteractions} below).

\subsection{Effects of interaction and disorder on the DOS of single TLSs}
\label{Shift of the DOS}

\begin{figure}[ht]
	\centering
	\includegraphics[scale=0.7]{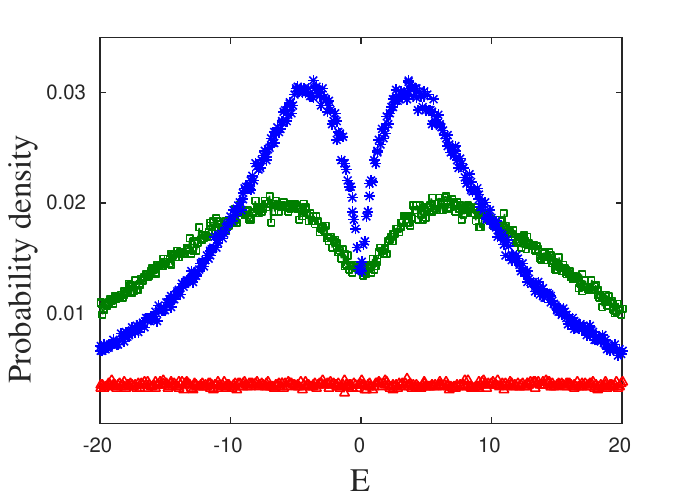}
	 \caption{(Color online) TLS DOS for different disorder values.  $W=1$ (blue stars), $W=10$ (green squares), and $W=100$ (red triangles) for $W/T=10$, constant interactions $J=1$ and $N=1000$. $r_{nn}=1$ as in all our calculations.}
	  \label{fig: tls dos DiffDisorder}
\end{figure}

\begin{figure}[ht]
	\centering
	\includegraphics[scale=0.7]{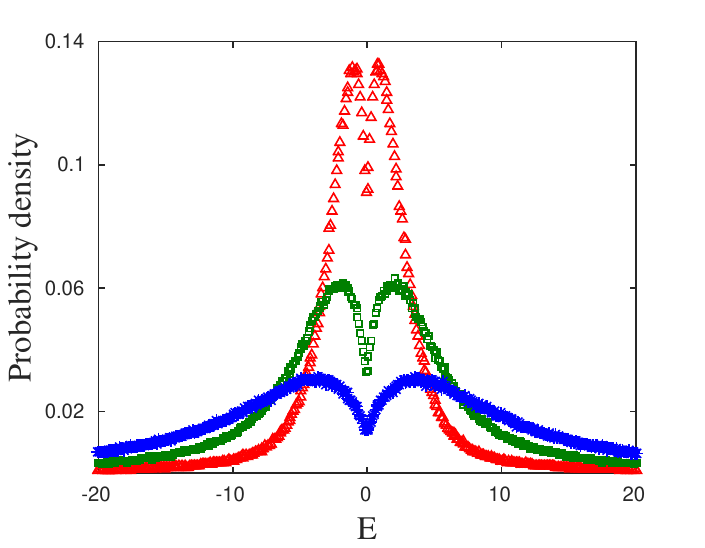}
	 \caption{(Color online) TLS DOS for different interaction and disorder values. $W/J=1$ (blue asterisks), $W/J=3.1$ (green squares), and $W/J=9.8$ (red triangles), keeping $W^2+J^2=2$. $N=1000$ and $T=0.1$. The structure of the DOS including the gap and the peaks around it is getting narrower and higher for larger values of $W/J$.}	
\label{fig: tls dos DiffInt}
\end{figure}

We present two schemes:
\begin{enumerate}
\item{Varying the disorder ($W$) for constant interactions ($J$) and constant $W/T$ ratio. For increasing disorder the DOS broadens and the gap diminishes (see Fig.~\ref{fig: tls dos DiffDisorder}). We note that a similar broadening is obtained for varying the disorder $W$ while keeping $T=0.1 J$ constant.}
\item{Varying the the ratio $J/W$ while holding the sum of the variances constant ,$W^2+J^2=2$. This is done in order to change the strength of the interactions while not significantly affecting of the energy variance,
    \begin{equation}
\label{TLS Energies variance}
\left<E_i^2\right> \approx W^2+\left<\Delta^2_{0i}\right> + \sum_j\frac{U_0^2}{r^6_{ij}}\left<S^2_j\right>.
\end{equation}
With the increase of the interaction strength (and decreasing of $W$) the overall effect is a broadening of the DOS and a deepening of the gap (see Fig.~\ref{fig: tls dos DiffInt}). For a similar scheme where only the interactions parameter is increased a greater broadening is obtained since $W$ is kept constant.}
\end{enumerate}


The study of the DOS of the TLS-glass is of interest by itself, but also in view of our interest in the dynamics of the TLS-glass. As can be inferred from Eq.~(\ref{relaxation rate}), the dynamics of the TLS-glass is strongly affected by the distribution of the single-TLS DOS. In fact, for a given realization of TLSs and constant temperature, there is a unique correspondence between the distribution of TLS energies and their dynamics. Thus, the change in DOS as function of varying disorder and interactions is a predictor of the change in the dynamics of the TLS-glass.
In Fig.~\ref{fig: tls dos DiffDisorder} and Fig.~\ref{fig: tls dos DiffInt} we plot the single-TLS DOS as a function of varying disorder and interaction according to the protocols described above (see also figure caption). We find that for larger $W$ or $J$ the width of the DOS increases. Also, for larger ratio $J/W$ the depth of the gap increases as expected. Note that in the disorder variation scheme, large values of disorder are taken in order to obtain a large enough qualitative effect in the rates distribution (plotted in Sec.~\ref{Shift of the distribution of relaxation rates} below).

\subsection{Shift of the distribution of relaxation rates}
\label{Shift of the distribution of relaxation rates}

\begin{figure}[ht]
\centering
\begin{subfigure}
  \centering
  \includegraphics[scale=0.7]{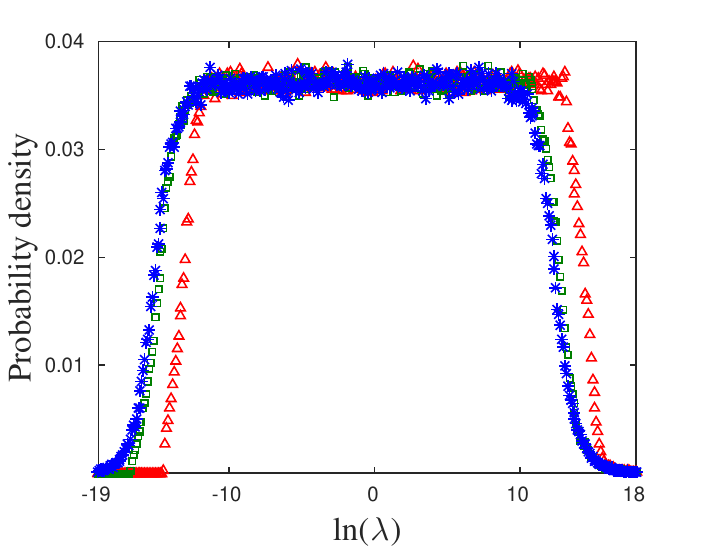}
\end{subfigure}
\begin{subfigure}
  \centering
  \includegraphics[scale=0.7]{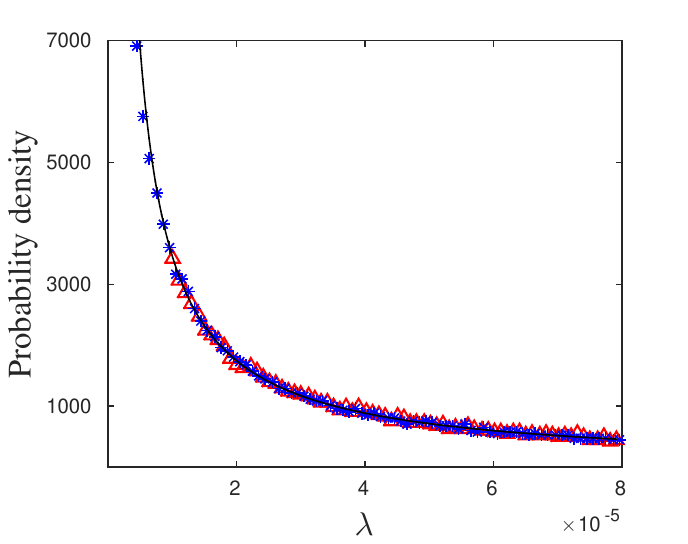}
\end{subfigure}
\caption{(Color online) TG decay rate distribution for different disorder values. $W=1$ (blue stars), $W=10$ (green squares), and $W=100$ (red triangles). $W/T=10$, $J=1$, and $N=1000$. (a) Rate distributions in log scale. (b) Rate distributions in normal scale. The values for $W=10$ are similar to those of $W=1$ and are therefore discarded.}
\label{fig: TLS Rates DiffDisorder}
\end{figure}
\begin{figure}[ht]
\centering
\begin{subfigure}
  \centering
  \includegraphics[scale=0.7]{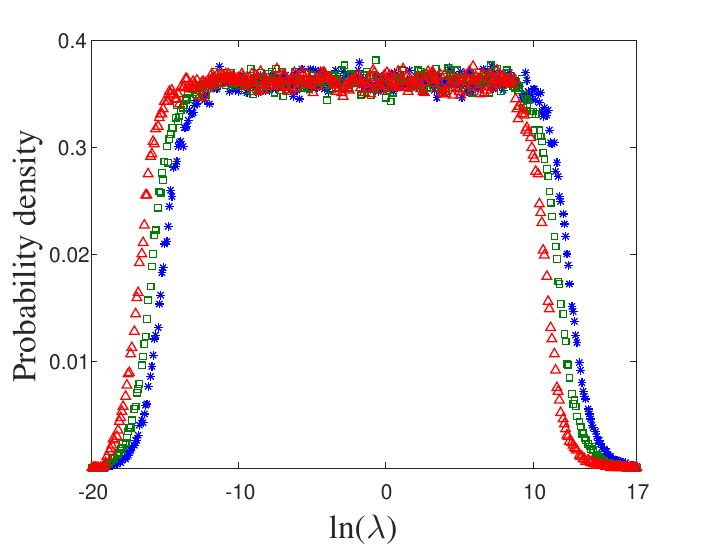}
\end{subfigure}
\begin{subfigure}
  \centering
  \includegraphics[scale=0.7]{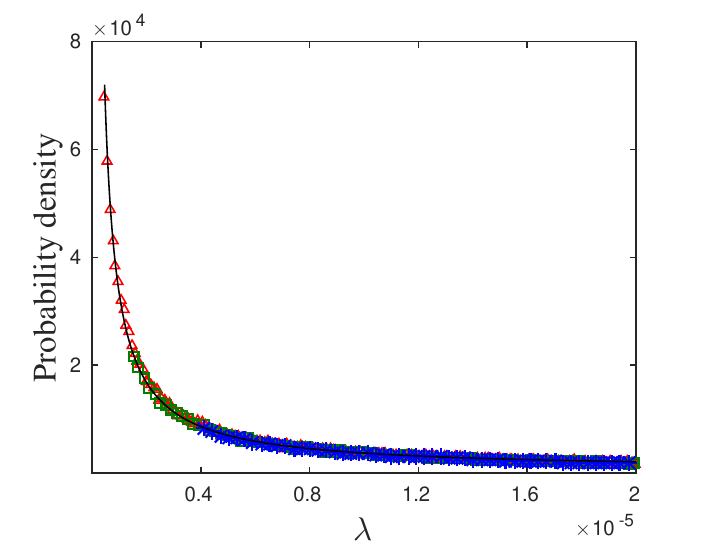}
\end{subfigure}
\caption{(Color online) TG decay rate distribution for different interaction and disorder values. $W/J = 1$ (blue asterisks), $W/J = 3.1$ (green squares), and $W/J = 9.8$ (red squares). The rest of the parameters are the same as in Fig.~\ref{fig: tls dos DiffInt}. (a) Rate distributions in log scale. (b) Rate distributions in normal scale.}
\label{fig: TLS Rates DiffInt}
\end{figure}
The distributions of relaxation rates [Eq.~(\ref{relaxation rate})] for both of the schemes presented in Sec.~\ref{Shift of the DOS} are plotted in Fig.~\ref{fig: TLS Rates DiffDisorder} for the variation of $W$ and in Fig.~\ref{fig: TLS Rates DiffInt} for the variation of $W/J$. As can be seen, for increasing $W$ or $J/W$ the rates distributions are shifted to higher values on the same $1/\lambda$ curve. This is a consequence of the shift of the lower cutoff with the variation of parameters. In turn, the lower cutoff represents TLSs which also have, in addition to small tunneling amplitude $\Delta_0$, small excitation energy. The number of such TLSs diminishes with the deepening of the gap and the enhancement of the variance of the DOS, leading to faster dynamics. Note that the upper cutoff is held fixed in the normal scale plots. This is due to the fact that the rates distribution extends over many orders of magnitude which are not relevant to the relaxation of the system at long time scales, i.e., $t \sim\lambda_{min}^{-1}$.

\section{Comparison to the electron-glass model}
\label{Comparison to the EG model}

In this section we consider the electron-glass (EG) model and compare its equilibrium and dynamical properties to the results of the TG model shown in Sec.~\ref{Results}. In Sec.~\ref{EG model and dynamics} we review the results of Amir {\it et al.} \cite{Ariel'sPaper}. We present the EG Hamiltonian, its equilibrium mean-field energies and the logarithmic relaxation which results from the $\frac{1}{|\lambda|}$ distribution of rates. In Sec.~\ref{EG model results} we address the effects of the disorder and interactions on the relaxation {to facilitate comparison between the TG and EG models, and add in this subsection a discussion of the effects of system size}. In Sec.~\ref{Structural comparison} we present additional similarities and differences which originate from the basic structure of the EG and TG models.

\subsection{The electron-glass: Model and dynamics}
\label{EG model and dynamics}

\begin{figure}[ht]
	\centering
	\includegraphics[scale=0.7]{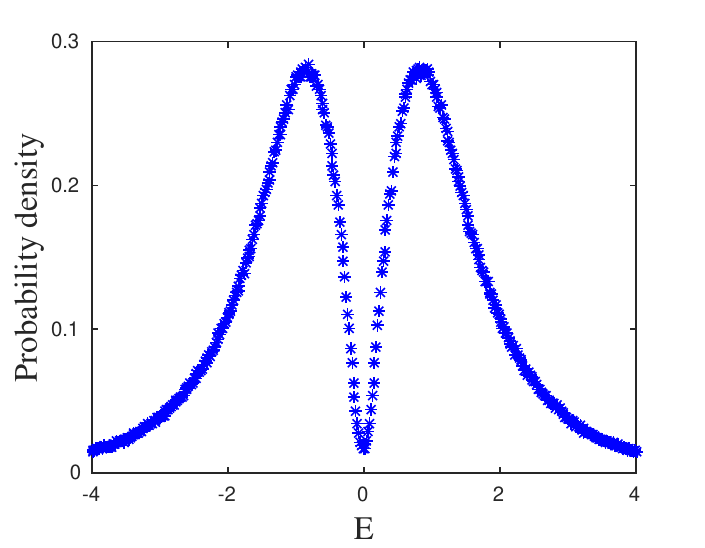}
	\caption{(Color online) EG DOS. The normalized histogram of site energies $E_i$ for half-filling and $N=10000$ sites \cite{Ariel'sPaper}. The energies $\epsilon_i$ are uniformly distributed in the interval $[-\frac{W}{2}, \frac{W}{2}]$, and  $\frac{e^2}{r_{nn}T}=20$. $W=1$ and $E_i$ are taken in units of interaction at average nearest-neighbor distance $J=\frac{e^2}{r_{nn}}$, where $r_{nn}$ is the average nearest-neighbor distance. The sites are distributed uniformly on a square with periodic boundary condition and averaged over 300 realizations.}
\label{fig: EG dos}
\end{figure}

The electron-glass (EG) system is composed of $N$ localized electronic states with random energies and $M < N$ electrons interacting via the unscreened Coulomb interaction. The electron-phonon coupling induces inter-site electron transitions. Since the Hubbard energy is assumed to be much greater than the energy scale of the system, only single occupation at each site is allowed. The exchange interaction is assumed to be much smaller than the Coulomb interaction, resulting in spinless electrons. Accordingly, the Hamiltonian of the EG system is \cite{BurinElectronGlass,2011ArielReview,pollak2012}
\begin{equation}
\label{Full EG model}
\mathcal{H}_{EG} = \sum^N_{i=1} \epsilon_i\left(n_i-K\right) + \sum^N_{i=1} \sum_{j>i} \frac{e^2}{r_{ij}}\left(n_i-K\right)\left(n_j-K\right)
\end{equation}
where $\epsilon_i$ are the random site energies of the system in the absence of interactions, $\frac{e^2}{r_{ij}}$ is the Coulomb interaction between the electrons at sites $i$ and $j$, $K=\frac{M}{N}$ is the background charge and $n_i, n_j \in [0,1]$ are site occupations. The sites are distributed uniformly in a square.
In equilibrium, the site occupations obey the Fermi-Dirac statistics, $n^0_i=\left(e^{E_i/T}+1\right)^{-1}$, and accordingly the self-consistent equations are
\begin{equation}
\label{eq: MF Electron glass equilibrium}
E_i =\epsilon_i - \frac{1}{2} \sum_{j\neq i} \frac{e^2}{r_{ij}}\tanh\left(\frac{E_j}{2T}\right) \, ,
\end{equation}
where $E_i$ are the MF energy of site $i$ and Boltzmann constant is set to unity.
The DOS obtained from the self-consistent Eq.~(\ref{eq: MF Electron glass equilibrium}) shows a gap around the chemical potential, known as the Coulomb-gap, first predicted by Efros and Shklovskii \cite{EfrosShklovskiiCoulombGapT=0}. Starting from randomly distributed values in each realization, the MF energies are found by an iterative procedure introduced by Grunewald {\it et al.} \cite{NumericalMethodForDOS}. The numerical solution in 2D gives a linear density of states for low energies\cite{Ariel'sPaper} (see also Fig.~\ref{fig: EG dos}). As can be seen finite temperature introduces a finite DOS at correspondingly low energies. Similarly to the DOS of the TG model, for large enough temperature the gap disappears completely \cite{DaviesCoulombGapT>0, LevinCoulombGapT>0, PikusCoulombGapT>0}.

\begin{figure}[ht]
\centering
\begin{subfigure}
  \centering
  \includegraphics[scale=0.7]{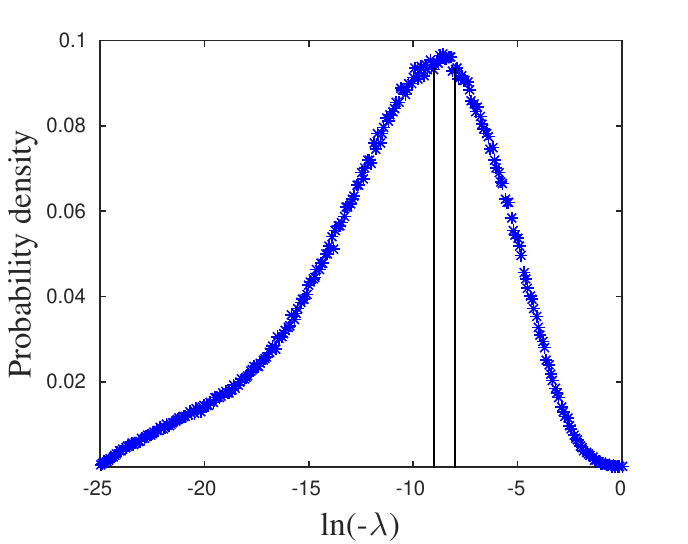}
\end{subfigure}
\begin{subfigure}
  \centering
  \includegraphics[scale=0.7]{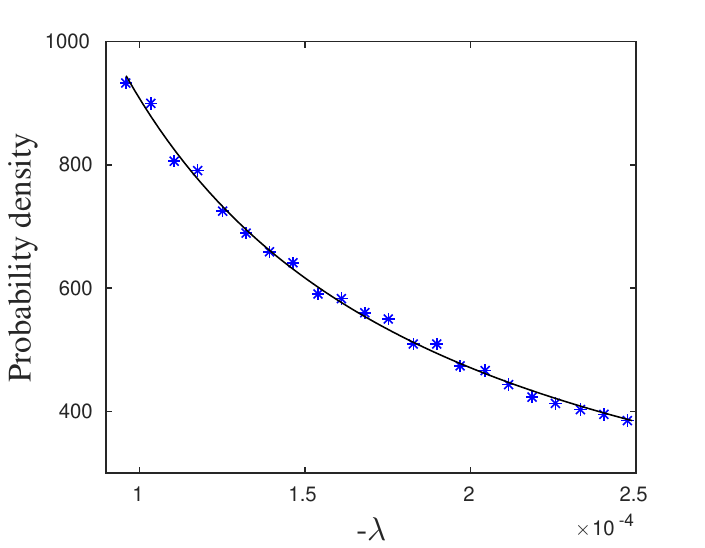}
\end{subfigure}
\caption{(Color online) Decay rate distribution for the electron-glass. Normalized histograms of the real part of the decay rates (originally done elsewhere \cite{Ariel'sPaper}), obtained by numerical diagonalization of the rate matrix $A_{ij}$ given in Eq.~(\ref{Rate matrix}), while neglecting the direct interactions term. $N=1000$, $\frac{e^2}{r_{nn}T}=10$, and $\frac{r_{nn}}{\xi}=10$. The disorder energy and density of sites are the same as in Fig.~\ref{fig: EG dos}. The graph is averaged over 1000 realizations.
	 (a) Rate distribution in log scale. The cutoff values are taken around the plateau region. (b) Rate distribution in normal scale with a $\frac{1}{\lambda}$ fit. The region of the plot is determined by the cutoffs as seen in the log plot.}
\label{fig: EG 1/lambda}
\end{figure}

The dynamics of the average electronic occupations is calculated using the Pauli-rate equation (\ref{eq: Master-Equation general}) with Miller and Abrahams transition rates \cite{Miller-Abrahams-Transitions}
\begin{equation}
\label{eq: Miller-Abrahams EG}
\gamma_{ij} =
    \Gamma^0_{ij} n_i\left(1-n_j\right)e^{-r_{ij}/\xi}\left[N(|\Delta E|)+\Theta(|\Delta E|)\right] \, .
\end{equation}
Here $\Theta$ is a step function, $N=\left(e^{|\Delta E|/T}-1\right)^{-1}$ is the phonon occupation, $\Delta E = E_i-E_j$ and $\xi$ is the localization length of the electron. The prefactor is $\Gamma^0_{ij} \simeq \frac{2\pi}{\hbar}|M_q|^2\nu$, where $M_q$ is the strength of the electron-phonon interaction and $\nu$ is the phonon density of states. Since we are interested in a qualitative description of the dynamics, the rates will be presented in units of $\Gamma^0_{ij}$.
The linearized rate equation for small deviation around equilibrium values is given by \cite{Ariel'sPaper}
\begin{equation}
\label{eq: linearized rate equation EG}
\frac{d \delta n_i}{dt} = \sum_j A_{ij}\delta n_j \,
\end{equation}
where the rate coefficients matrix take the form
\begin{equation}
\label{Rate matrix}
A_{ij} =
\begin{cases}
\frac{\gamma^0_{ij}}{n^0_j \left(1-n^0_j\right)}-\sum_{k\neq j,i}\frac{e^2 \gamma^0_{ik}}{T} \left(\frac{1}{r_{ij}}-\frac{1}{r_{jk}}\right) &; \; i \neq j\\
-\sum_k A_{kj} &; \; i=j \, .
\end{cases}
\end{equation}
The superscript $"0"$ indicates equilibrium values. The diagonal elements $A_{ii}$ are dictated by the requirement of particle number conservation, $\sum_i A_{ij}=0$. Neglecting the second term of the off-diagonal element of $A_{ij}$, the top line in Eq.(\ref{Rate matrix}) (the electron-electron interaction term has been shown to be insignificant at low temperatures \cite{Ariel'sPaper, PhysRevB.80.245214}), one obtains the $\frac{1}{|\lambda|}$ distribution of relaxation rates \cite{Ariel'sPaper, PhysRevLett.105.070601}; see also Fig.~\ref{fig: EG 1/lambda}. Notice that according to the definition of Eq.~(\ref{Rate matrix}) the rates will turn out to be negative.
Solving the linearized rate equation and going through the steps shown in Sec.~\ref{Dynamics}, the obtained total relaxation of the EG systems for times $\frac{1}{\lambda_{max}} < t < \frac{1}{\lambda_{min}}$ and $p(\lambda)= \frac{1}{\lambda}$ rate distribution is \citep{Ariel'sPaper}:
\begin{equation}
\label{eq: |n| as an integral}
|\bm{\delta n}| \simeq c\int_{\lambda_{min}}^{\lambda_{max}} \frac{e^{-\lambda t}}{\lambda} d\lambda \simeq -c[\gamma_{E}+log(\lambda_{min}t)]
\end{equation}
where the assumption is that the rate matrix eigenvectors are excited roughly with a uniform probability $c(\lambda)\simeq c$ except for the eigenvector associated with the zero eigenvalue which can not be excited since the total particle number is conserved.

\subsection{The effect of interactions and disorder in the EG model}
\label{EG model results}

\begin{figure}[ht]
	\centering
	\includegraphics[scale=0.7]{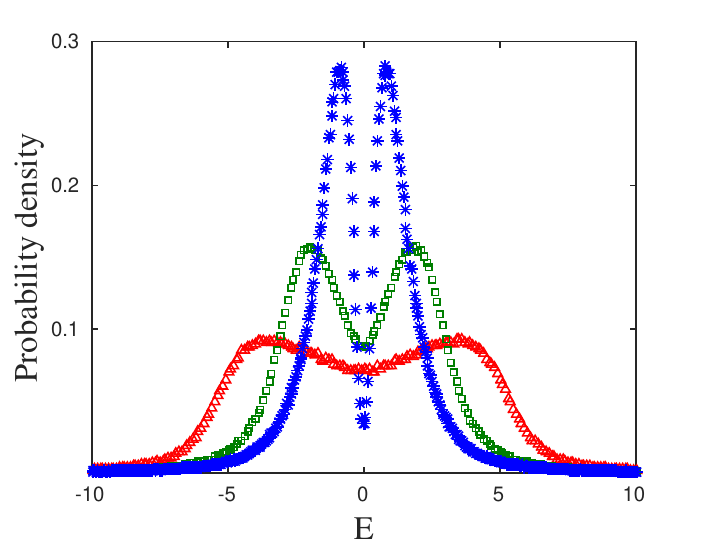}
	\caption{(Color online) EG DOS for different disorder values. $W=1$ (blue stars), $W=5$ (green squares) and $W=10$ (red triangles) for $W/T=10$, constant interactions $J=1$, and $N=1000$. The density of sites is set as in Fig.~\ref{fig: EG dos}.}
\label{fig: EGDosDiffDisorder}
\end{figure}
\begin{figure}[ht]
\centering
\begin{subfigure}
  \centering
  \includegraphics[scale=0.7]{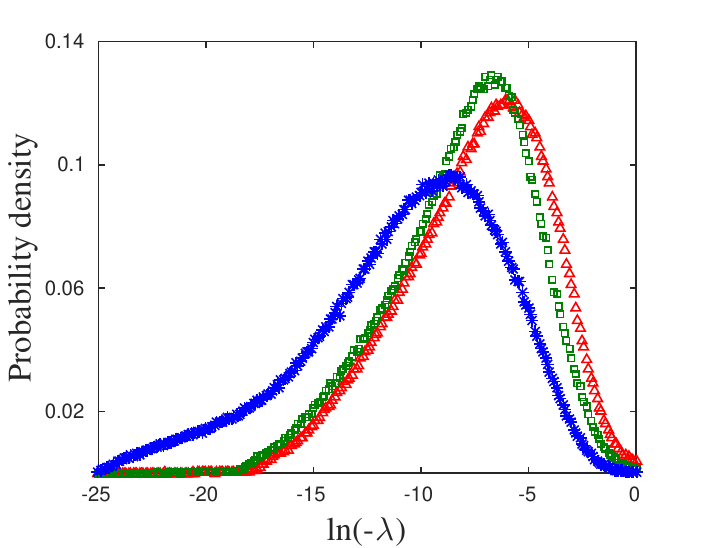}
\end{subfigure}
\begin{subfigure}
  \centering
  \includegraphics[scale=0.7]{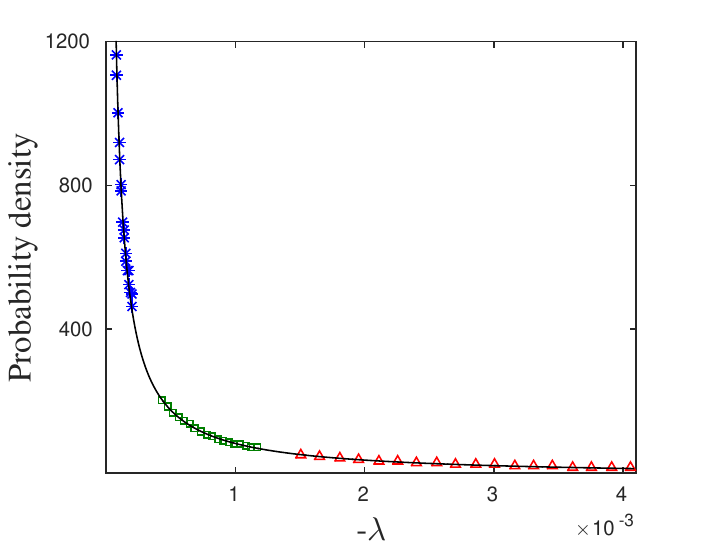}
\end{subfigure}
\caption{(Color online) EG decay rate distribution as given by the distribution of eigenvalues of the rate matrix in Eq.~(\ref{eq: linearized rate equation EG})\cite{Ariel'sPaper}, for different disorder values. The parameters are the same as in Fig.~\ref{fig: EGDosDiffDisorder}. (a) Rate distributions in log scale. (b) Rate distributions in linear scale.}
\label{fig: EG Rate diff disorder}
\end{figure}
\begin{figure}[ht]
	\centering
	\includegraphics[scale=0.7]{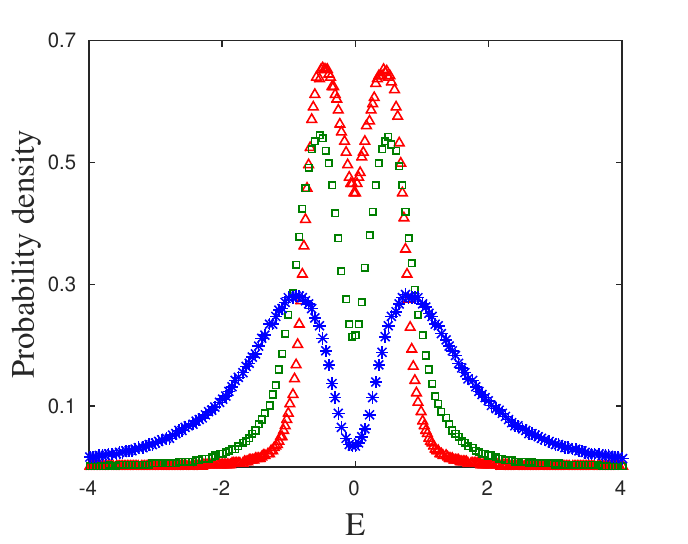}
	\caption{(Color online) EG DOS for different interaction and disorder values. $W/J=1$ (blue stars), $W/J=\sqrt{17}$ (green squares) and $W/J=\sqrt{97}$ (red triangles) for constant sum of variances $W^2+J^2=2$, temperature $T=0.1$, and $N=1000$. The density of sites is set as in Fig~.\ref{fig: EG dos}.}
\label{fig: EGDosDiffInteractions}
\end{figure}
\begin{figure}[ht]
\centering
\begin{subfigure}
  \centering
  \includegraphics[scale=0.7]{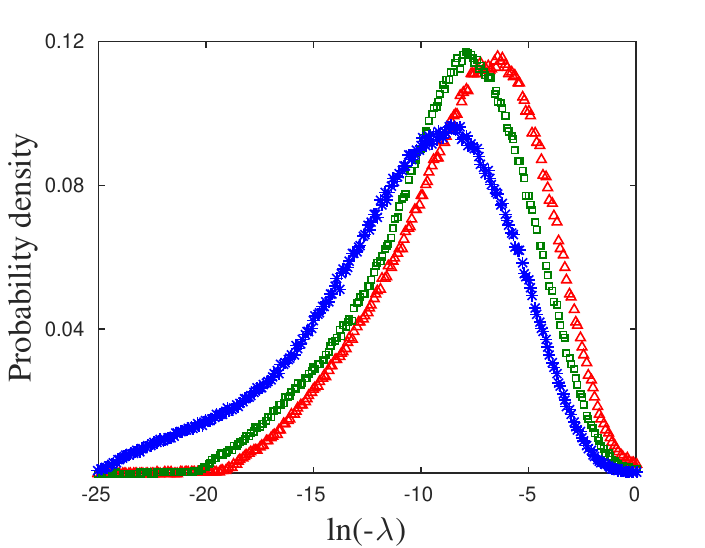}
\end{subfigure}
\begin{subfigure}
  \centering
  \includegraphics[scale=0.7]{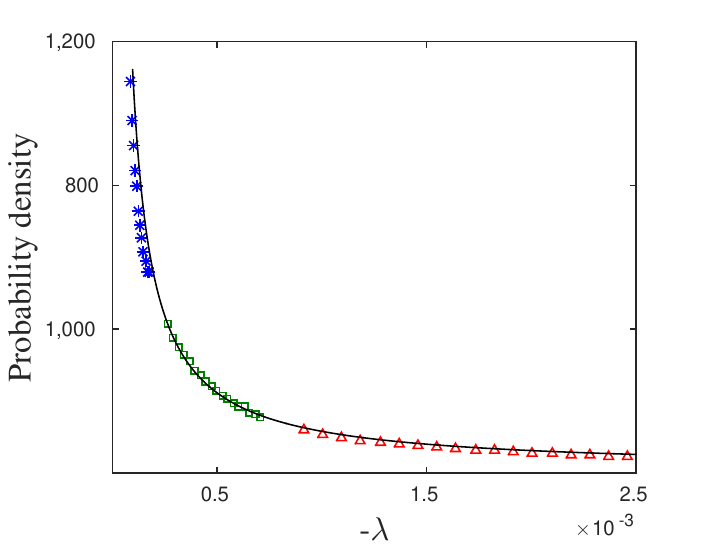}
\end{subfigure}
\caption{(Color online) EG decay rate distribution as given by the distribution of eigenvalues of the rate matrix in Eq.~(\ref{eq: linearized rate equation EG})\cite{Ariel'sPaper}, for different values of the parameter $J/W$. The parameters are the same as in Fig.~\ref{fig: EGDosDiffInteractions}. (a) Rate distribution in log scale. (b) Rate distributions in linear scale.}
\label{fig: EG Rate diffInteractions}
\end{figure}

To compare between the TG model and the EG model we perform the same parameter-varying schemes for the electron-glass model as presented in the previous section for the TG model (i.e., varying $W$ and $J/W$; see Sec.~\ref{Shift of the DOS}) and study how the DOS and rates are affected. This is done by studying the typical change of the rate matrix element, Eq.~(\ref{Rate matrix}), without the electron-electron interaction term \cite{Ariel'sPaper, PhysRevB.80.245214}, and use it as a measure for the shift of the rates.
Note that the plateau region in the rate log plots of the EG model is narrower than in the TG model allowing us conveniently to take also the upper cutoff.
Figures \ref{fig: EGDosDiffDisorder} and \ref{fig: EG Rate diff disorder} show respectively that for increasing disorder the DOS broadens while the gap diminishes (although the enhancement of the DOS near zero energy is a direct consequence of the variation of temperature in that scheme) leading to a shift of the rates to higher values, even though the disorder is stronger.
Figures \ref{fig: EGDosDiffInteractions} and \ref{fig: EG Rate diffInteractions} show respectively how for increasing interactions the DOS broadens and the gap deepens, and at the same time a shift of the rates to lower values, opposite to the effect of interactions on the TG relaxation rates.
Unlike the case for the TLS-glass, for the EG the connection between the single particle DOS and the relaxation rate distribution is indirect. The DOS affects the $N^2-N$ hopping rates $A_{ij}$, which constitute the matrix whose $N$ eigenvalues are the relaxation rates. Still, some intuition may be obtained by considering the dependence of $A_{ij}$ on the energy difference between the sites $i,j$ for low temperatures:
\begin{equation}
\label{eq: Approximated matrix element}
A_{ij} \sim N(\Delta E) \sim
\begin{cases}
	1 					 & , \Delta E>T, \; \Delta E<0\\
	e^{-\Delta E/T} 	 & , \Delta E>T, \; \Delta E>0\\
	\frac{T}{|\Delta E|} & , |\Delta E|<T
\end{cases}.
\end{equation}
Narrower gaps and a larger DOS at low energies enhance the weight of small energy differences between near neighbor sites, which in turn leads to faster relaxation.

Finally, it is worth mentioning the results obtained for changing the system size. Both for the EG and TG models we found numerically that increasing the number of sites (while keeping a constant density) shifts the rate distribution to lower values. Specifically for the TG model, we found the shift to be negligible whereas for the EG the effect is more pronounced. It turns out that the shift in the EG model is a finite-size effect that originates from the statistics of the exponential distance matrix ($e^{-r_{ij}/\xi}$) rather than from the dependence on the interactions; see App.~\ref{Appendix_1}. The fact that the interactions have a negligible contribution to the change in dynamics as the system size is enhanced, in both models, suggests that the relaxation modes are local in nature.

Table~\ref{TableEGTG} summarizes the effects of disorder, interactions, and system size on the dynamics of the EG and TG models.
\begin{table}[ht]
    \begin{tabular}{| p{2.3cm} | p{1.9cm} | p{1.9cm} | p{1.9cm} |}
    \hline
    Model/Quantity & Disorder & Interaction & System size\\ \hline
    EG & $+$ & $-$ & $-$ \\ \hline
    TG & $+$ & $+$ & $-$ \\
    \hline
    \end{tabular}
    \caption{(Color online) Comparison between the relaxation dynamics of the EG and the TG models for increasing disorder $(W)$, interactions $(J)$, and system size $(N)$. The $(+)$ and $(-)$ signs indicate faster and slower relaxation respectively. Note that for changing the system size, the dynamics has a weak dependence on the mean-field energies (and thus on the interaction $J$) in both models, which implies that the relaxation modes are local.}
    \label{TableEGTG}
\end{table}

\subsection{Structural comparison of the EG and TG models}
\label{Structural comparison}
In this section we compare the formal solutions of the EG and TG models.
First, comparing Eqs.~(\ref{relaxation rate}), (\ref{Full linearzed rate equation})and Eqs.~(\ref{Rate matrix}), (\ref{eq: linearized rate equation EG}) we see that the interaction term of the EG rate equation includes also an interaction and transition with a third site whereas the second term in the TG rate equation does not. This difference stems from the fact that transitions are allowed only within pairs of states (TLSs act as dimers). The interactions are then between two dimers, whereas in the EG model the interactions are between single site occupations. This leads to the notion that under a certain condition one may obtain the mathematical structure of the TG model from the given EG model. This happens when the distance between next nearest-neighbor ($r_{nnn}$) is sufficiently larger than the nearest-neighbor distance ($r_{nn}$), i.e., $r_{nnn}-r_{nn} > 2\xi$.

Also, in both models the $\frac{1}{|\lambda|}$ distribution of relaxation rates leading to logarithmic relaxation is a result of the wide and rather homogeneous distribution of an exponent, i.e., $A_{ij} \propto e^{-r_{ij}/\xi}$ in Eq.~(\ref{Rate matrix}) in the EG model and $\Delta_0 \propto e^{-\Lambda}$ in the TG model. However, the range in which the $\frac{1}{|\lambda|}$ form is satisfied is much wider in the TG model (Fig.~\ref{fig: TLS Rate}) than in the EG's (Fig.~\ref{fig: EG 1/lambda}). This is a result of the exponent in the TLS glass model being chosen as homogeneous over a large regime, whereas in the EG model the tunneling amplitude is dictated by the distribution of nearest-neighbor distances, which is narrower. Last, the different dependence of the rates on the mean-field energies in the two models leads to the different consequences of varying the interactions, disorder, and system size on the dynamics of the two models.

\section{Summary and conclusions}
\label{conclusions}

In this work we examine thermodynamic and dynamic properties of the TLS-glass, modeled by the transverse field Ising model with random $1/r^3$ interactions and random local fields. Using mean field approximation, we first rederive the single particle DOS for this model, and then derive the dynamics of its relaxation to equilibrium. Similarly to the electron glass model, we find $1/\lambda$ distribution of relaxation rates, leading to logarithmic time relaxation and known memory effects in such models \cite{AgingEffectsVaknin,SlowRelaxationsAmir}.
We further find that increasing the disorder shifts the rate distribution to higher values, similarly to what was observed for the electron glass\cite{Ariel'sPaper}, but increasing the interactions shifts the rate distribution to higher values while in the EG model this results in a shift to lower values\cite{Ariel'sPaper}. This suggests that the effect of interactions on glass dynamics is system dependent. Finally, we show that the interactions have a negligible effect on the rate distribution for changing the system size at constant site density, which implies that the relaxation modes are localized.

Given the complexity of the EG and TG models we use the MF approximation which simplifies the calculation. It would be of interest to check our results for the dynamics of the system, some of them unexpected, against more exact numerical methods such as Monte Carlo simulations or exact diagonalization of finite systems.

\begin{acknowledgments}
We would like to thank Alexander Burin, Doron Cohen, and Zvi Ovadyahu for useful discussions. This work was supported by the Israel Science Foundation (Grant No. 821/14) and by the German-Israeli Foundation (GIF Grant No. 1183/2011).
\end{acknowledgments}

\appendix

\section{Finite-size effect in the EG model}
\label{Appendix_1}
In this appendix we give a qualitative argument that explains the shift of the rate distribution caused by changing the system size in the EG model.
\begin{figure}[]
	\centering
	\includegraphics[scale=0.7]{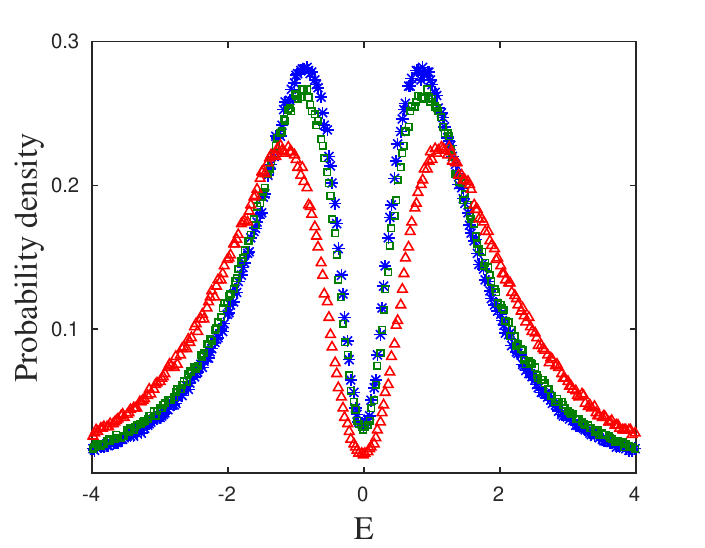}
	\caption{(Color online) EG DOS for different system sizes. The number of sites $N$ is varied, $10$ (red triangles), $100$ (green squares), $1000$ (blue asterisks). $\frac{e^2}{r_{nn}T}=10$. The disorder and density of sites are the same as in Fig.~\ref{fig: EG dos}. Notice how the DOS is wider and the gap is shorter for smaller number of sites.
	 The same quantitative behavior can be seen for changing the interactions strength to smaller values.}
\label{fig: EG DOS DiffSiteNum}
\end{figure}
\begin{figure}[]
\centering
\begin{subfigure}[]
  \centering
  \includegraphics[scale=0.7]{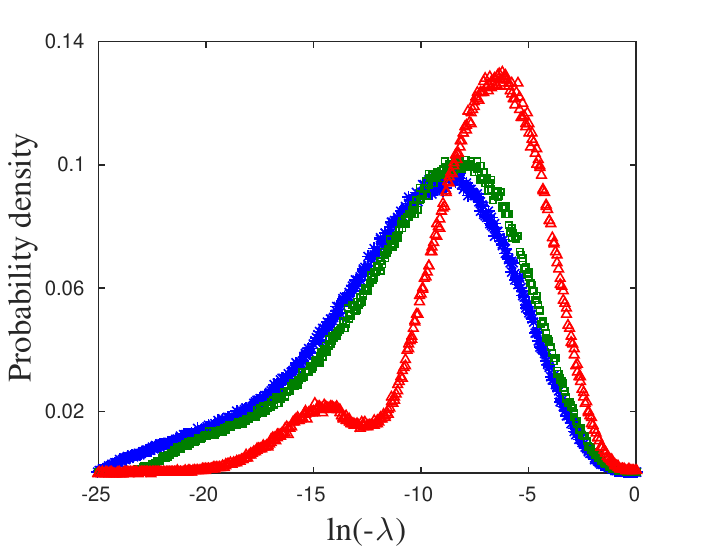}
\end{subfigure}
\begin{subfigure}[]
  \centering
  \includegraphics[scale=0.7]{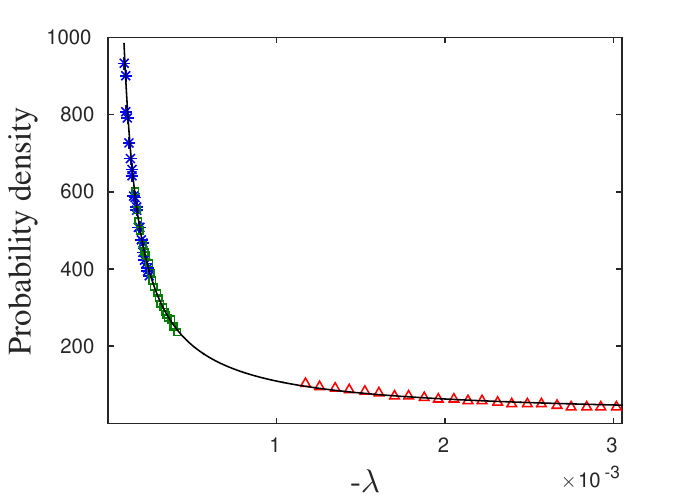}
\end{subfigure}
\caption{(Color online) EG decay rate distribution as given by the distribution of eigenvalues of the rate matrix in Eq.~(\ref{eq: linearized rate equation EG}), for different system sizes. $N=10$ (red triangles), $N=100$ (green squares), and $N=1000$ (blue asterisks). Localization length $\xi=0.1$. Disorder energy $W$ and density of sites are the same as in Fig.~\ref{fig: EG dos}. The solid line is a fit of $1/x$ curve. (a) Rate distributions in the natural log scale. (b) Rate distributions in linear scale.}
\label{fig: EG Rate diffSiteNum}
\end{figure}
\begin{figure}[]
\centering
\begin{subfigure}[]
  \centering
  \includegraphics[scale=0.7]{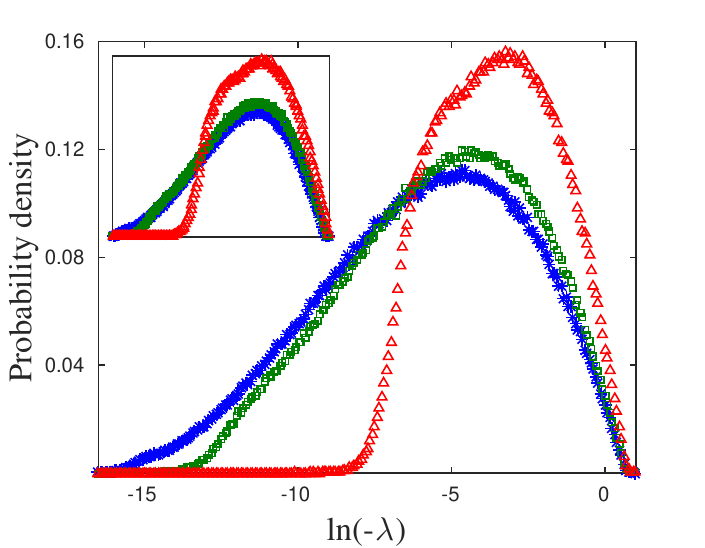}
\end{subfigure}
\begin{subfigure}[]
  \centering
  \includegraphics[scale=0.7]{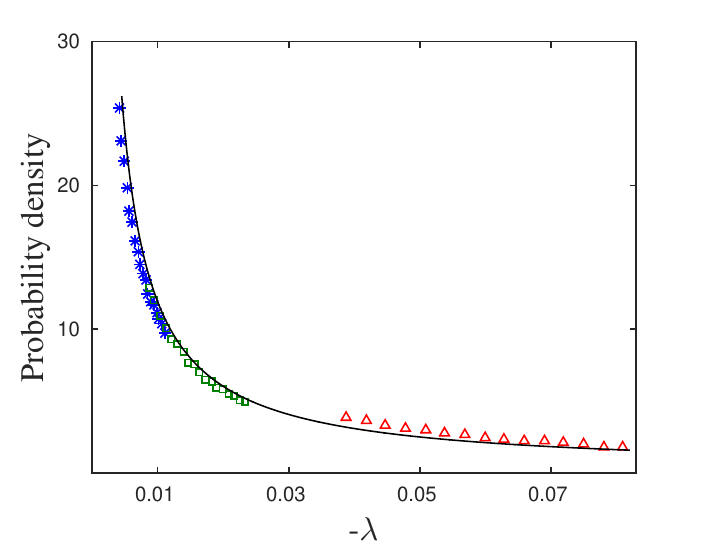}
\end{subfigure}
\caption{(Color online) EG rate distribution as given by the eigenvalue distribution of the matrix $A_{ij}=e^{-r_{ij}/\xi}$ (without energy dependence), plotted for different system sizes. $N=1000$ (blue stars), $N=100$ (green squares), and $N=10$ (red triangles). The parameters are the same as in Fig.~\ref{fig: EG Rate diffSiteNum}. (a) Rate distributions in the natural log scale. The inset shows how the peaks almost coincide after scaling $\xi$ with the prefactor given in Eq.~(\ref{nearest neighbors distance log rate distribution}). (b) Rate distributions in linear scale.}
\label{fig: EGRatesChangingNOnlyExp}
\end{figure}

In Fig.~\ref{fig: EG DOS DiffSiteNum} the EG DOS is plotted for different system sizes, showing a narrowing of the gap for increasing size. In Fig.~\ref{fig: EG Rate diffSiteNum} the EG relaxation rate distribution is plotted showing a shift to lower values for increasing system size, which might seem counter intuitive given the behavior of the DOS. We show below how this shift is dominated by the tunneling term.

Figure \ref{fig: EGRatesChangingNOnlyExp} shows the rate distribution as given in Fig.~\ref{fig: EG Rate diffSiteNum} after excluding the interactions term in the rate matrix, i.e., $A_{ij} = e^{-r_{ij}/\xi}$. Comparing the two graphs it is evident that the shift of the rate distribution peaks remains qualitatively the same.

Let us now estimate this effect. For $r_{nn} \ll \xi$, the relaxation is dominated by tunneling of electrons to their near neighbor site. Given a linear size of the sample $L$ it can be shown that the distribution of nearest-neighbor distance for $r_{nn} \ll L$ is\cite{Ariel'sPaper}
\begin{equation}
\label{nearest neighbors distance distribution}
p(r) = \frac{V_d}{L^d}d(N-1)r^{d-1}e^{-V_d(N-1)(r/L)^d} \, ,
\end{equation}
where $r$ is the nearest-neighbor distance in the continuous limit, $d$ is the dimension, and $V_d$ is of order unity, e.g. $V_1 = 2$, $V_2 = \pi$. Substituting the average nearest-neighbor distance $\bar{r}=L^d/(N^{1/d}-1)$ for constant site density, as used in the numerical calculation, and changing to variable $x = \log(-\frac{\lambda}{2})=-\frac{r}{\xi}$\cite{Ariel'sPaper} we obtain for $d=2$:
\begin{equation}
\label{nearest neighbors distance log rate distribution}
p(x) = 2\pi \left(\frac{\xi}{\bar{r}}\right)^2 x \frac{\sqrt{N}+1}{\sqrt{N}-1} \exp {\left[-\pi \left(\frac{\xi}{\bar{r}} \right)^2 x^2 \frac{\sqrt{N}+1}{\sqrt{N}-1} \right]} \, .
\end{equation}
As can be seen from Eq.~\ref{nearest neighbors distance log rate distribution}, $p(x)$ shifts with system size ($N$) in the same qualitative manner as obtained numerically in Fig.~\ref{fig: EGRatesChangingNOnlyExp}. Furthermore, the inset shows the distributions when scaled according to Eq.~\ref{nearest neighbors distance log rate distribution} $x \rightarrow x\left(\frac{\sqrt{N}+1}{\sqrt{N}-1}\right)^{\frac{1}{2}}$. This suggests that within the mean-field approximation discussed in this work, the dominant cause for the slow down of relaxation is a finite-size effect that changes the effective average near neighbor distance.
In similarity to the EG, the TG also shows slowing down of the relaxation with increased system size, but the effect is much weaker (not shown). Note though that for the TG model the relaxation rates are dictated by the tunneling amplitudes $\Delta_{0i}$, which are distributed independently from the site distribution of the TLSs and therefore the relaxation rate distribution is not sensitive to finite-size effects.

\section{Interaction term in the rate equation}
\label{Appendix_2}

In this appendix we discuss the full linearized rate equation (Eq.~\ref{eq: rate equation TLS up to 1st order} with the addition of interaction terms)
\begin{equation}
\label{Full linearzed rate equation}
\frac{d\sigma_i}{dt} \simeq -\lambda_i \delta \sigma_i + \sum_{j\neq i} f_{ij} \delta \sigma_j \equiv \sum_j A_{ij} \delta \sigma_j
\end{equation}
and show that the second term can be neglected at low temperatures.
Here $f_{ij} = 2a_i \Delta^2_{0i} \Delta'_i N_i(N_i+1)\sigma^0_i \frac{u_{ij}}{Tr^3_{ij}}$ is the interaction prefactor and $n_i=\left(e^{\beta|E_i|}-1\right)^{-1}$, $\sigma^0_i=-\tanh\left(\frac{E_i}{2T}\right)$ are the phonon and pseudospin equilibrium occupations.

An estimate for the contribution of interaction term to the rate equation is given by the ratio
\begin{equation}
\label{2}
\left| f_{ij}/\lambda_i \right| = \left| \frac{\Delta'_i}{|E_i|\cosh^2\left(\frac{E_i}{2T}\right)} \frac{u_{ij}}{Tr_{ij}^3}\right| \lesssim \left|\frac{\Delta'_i}{|E_i|\cosh^2\left(\frac{E_i}{2T}\right)} \frac{J}{T}\right| \, .
\end{equation}
The second inequality represents an upper bound, where $i,j$ are nearest-neighbor, i.e.
$\frac{u_{ij}}{r^3_{ij}} = \frac{U_0}{r^3_{nn}} \equiv J$.
Furthermore, since both $\lambda_i$ and $f_{ij}$ are proportional to $\Delta^2_{0i}$, the lowest rates (eigenvalues of $A_{ij}$), which dictate the slow relaxation
of the system, have a tunneling amplitudes
$\Delta_{0i} \sim \Delta_{0_{min}}$.
Thus, one can approximate $\Delta'_i\approx E_i$ and substitute in Eq.~\ref{2}:
\begin{equation}
\label{3}
\begin{split}
&\max(|f_{ij}/\lambda_i|) =\\
&=\begin{cases}
    \frac{J}{T} \left[1+\frac{1}{2}\left(\frac{E_i}{2T}\right)^2\right]^{-1} \longrightarrow \frac{J}{T}  & , |E_i| \ll T\\
	0.1\frac{J}{T} < |f_{ij}/\lambda_i| < \frac{J}{T} & , |E_i| < 2T \\
	2 \frac{J}{T} e^{-|E_i|/T}   & , |E_i| > 2T
\end{cases}
\end{split}
\end{equation}
As can be seem from Eq.~\ref{3}, at low temperatures relative to disorder and interactions ($T \ll \sqrt{W^2+J^2}$) the typical TLS energy is larger then the temperature and $f_{ij}$ is exponentially suppressed.

Let us further consider two different limit cases:
(i) $T\sim W \ll J$. In this case TLSs with $f_{ij} \sim \lambda_i$ are rare since the temperature resides deep inside the dipole gap. (ii) $T\sim J \ll W$. In this case the gap is small but the DOS is flat and wide in comparison to the temperature scale, and the number of TLSs with energy smaller then the temperature
is $N(|E_i|<2T) \propto \frac{T}{W}\ll 1$.

\begin{figure}[]
	\centering
	\includegraphics[scale=0.4]{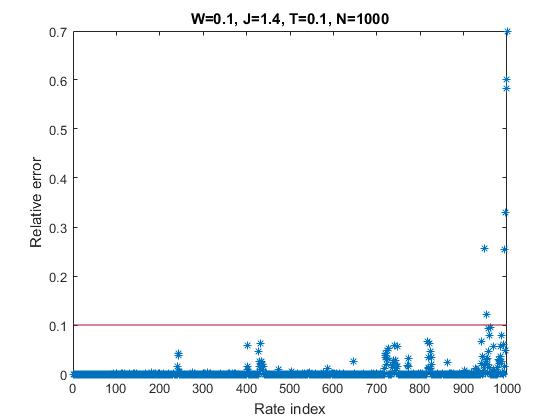}
	\caption{Relative error between the rates with and without the interaction term $f_{ij}$, for W=0.1, J=1.4, T=0.1, and N=1000.}
\label{fig1}
\end{figure}
\begin{figure}[]
	\centering
	\includegraphics[scale=0.4]{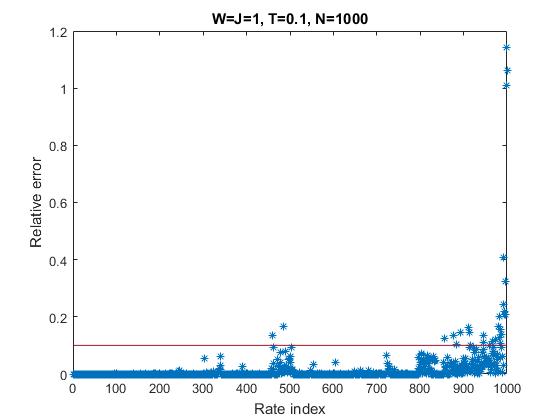}
	\caption{Relative error between the rates with and without the interaction term $f_{ij}$, W=J=1, T=0.1, and N=1000.}
\label{fig2}
\end{figure}
\begin{figure}[]
	\centering
	\includegraphics[scale=0.4]{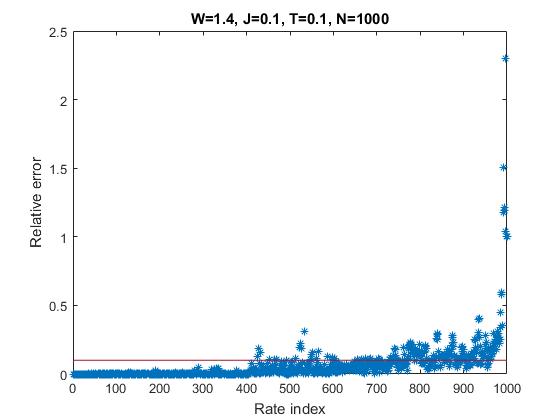}
	\caption{Relative error between the rates with and without the interaction term $f_{ij}$, W=1.4, J=0.1, T=0.1, and N=1000.}
\label{fig3}
\end{figure}
\begin{figure}[]
	\centering
	\includegraphics[scale=0.4]{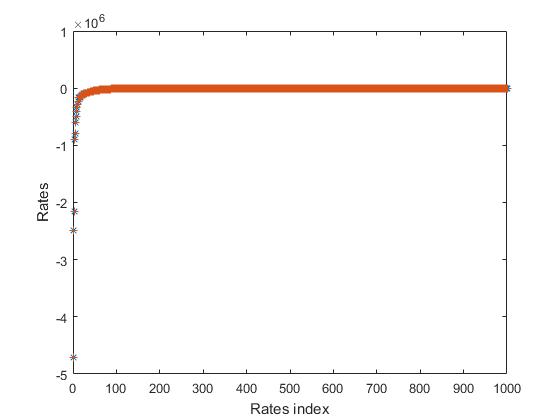}
	\caption{Typical graph of the rates with and without the interaction term for the three cases stated above. Blue (asterisk) for $f_{ij}=0$ and orange (plus) is for $f_{ij}\neq 0$.}
\label{fig4}
\end{figure}

To substantiate the conclusion that the interaction term in the rate equation has little effect on the dynamics of the system at low temperatures, we 
diagonalize numerically the rate matrix $A_{ij}$ in Eq.~(\ref{Full linearzed rate equation}) and compare with the results obtained neglecting the interaction term. 
The graphs presented below show 
the relative error, defined as $\left|\frac{\lambda'-\lambda}{\lambda'}\right|$, where $\lambda'$ ($\lambda$) are the rates including (excluding) interactions, 
for three cases: 
(i) $W \simeq 1.4, J \simeq 0.1$ (Fig~\ref{fig1}), (ii) $W \simeq 0.1, J \simeq 1.4$ (Fig~\ref{fig2}), (iii) $W = J = 1$ (Fig~\ref{fig3}). For all cases we numerically diagonalise a single realization of $A_{ij}$ for $N=1000$, $W^2+J^2 = 2$ and $T = 0.1$. 
In all cases low rates are negligibly affected by the interaction. In Fig~\ref{fig4} we present the rates for $W=J=1$ [case (i)] with and without interactions. Similar results were obtained for cases (ii) and (iii).

\bibliographystyle{apsrev4-1}
\bibliography{TLSsAndElectronsInAmorphousSolids}

\begin{thebibliography}{43}%
\makeatletter
\providecommand \@ifxundefined [1]{%
 \@ifx{#1\undefined}
}%
\providecommand \@ifnum [1]{%
 \ifnum #1\expandafter \@firstoftwo
 \else \expandafter \@secondoftwo
 \fi
}%
\providecommand \@ifx [1]{%
 \ifx #1\expandafter \@firstoftwo
 \else \expandafter \@secondoftwo
 \fi
}%
\providecommand \natexlab [1]{#1}%
\providecommand \enquote  [1]{``#1''}%
\providecommand \bibnamefont  [1]{#1}%
\providecommand \bibfnamefont [1]{#1}%
\providecommand \citenamefont [1]{#1}%
\providecommand \href@noop [0]{\@secondoftwo}%
\providecommand \href [0]{\begingroup \@sanitize@url \@href}%
\providecommand \@href[1]{\@@startlink{#1}\@@href}%
\providecommand \@@href[1]{\endgroup#1\@@endlink}%
\providecommand \@sanitize@url [0]{\catcode `\\12\catcode `\$12\catcode
  `\&12\catcode `\#12\catcode `\^12\catcode `\_12\catcode `\%12\relax}%
\providecommand \@@startlink[1]{}%
\providecommand \@@endlink[0]{}%
\providecommand \url  [0]{\begingroup\@sanitize@url \@url }%
\providecommand \@url [1]{\endgroup\@href {#1}{\urlprefix }}%
\providecommand \urlprefix  [0]{URL }%
\providecommand \Eprint [0]{\href }%
\providecommand \doibase [0]{http://dx.doi.org/}%
\providecommand \selectlanguage [0]{\@gobble}%
\providecommand \bibinfo  [0]{\@secondoftwo}%
\providecommand \bibfield  [0]{\@secondoftwo}%
\providecommand \translation [1]{[#1]}%
\providecommand \BibitemOpen [0]{}%
\providecommand \bibitemStop [0]{}%
\providecommand \bibitemNoStop [0]{.\EOS\space}%
\providecommand \EOS [0]{\spacefactor3000\relax}%
\providecommand \BibitemShut  [1]{\csname bibitem#1\endcsname}%
\let\auto@bib@innerbib\@empty
\bibitem [{\citenamefont {Zeller}\ and\ \citenamefont
  {Pohl}(1971)}]{Zeller&Pohl1971}%
  \BibitemOpen
  \bibfield  {author} {\bibinfo {author} {\bibfnamefont {R.~C.}\ \bibnamefont
  {Zeller}}\ and\ \bibinfo {author} {\bibfnamefont {R.~O.}\ \bibnamefont
  {Pohl}},\ }\href {\doibase 10.1103/PhysRevB.4.2029} {\bibfield  {journal}
  {\bibinfo  {journal} {Phys. Rev. B}\ }\textbf {\bibinfo {volume} {4}},\
  \bibinfo {pages} {2029} (\bibinfo {year} {1971})}\BibitemShut {NoStop}%
\bibitem [{\citenamefont {Anderson}\ \emph {et~al.}(1972)\citenamefont
  {Anderson}, \citenamefont {Halperin},\ and\ \citenamefont
  {Varma}}]{Anderson&Halperin&Varma}%
  \BibitemOpen
  \bibfield  {author} {\bibinfo {author} {\bibfnamefont {P.~W.}\ \bibnamefont
  {Anderson}}, \bibinfo {author} {\bibfnamefont {B.~I.}\ \bibnamefont
  {Halperin}}, \ and\ \bibinfo {author} {\bibfnamefont {C.~M.}\ \bibnamefont
  {Varma}},\ }\href@noop {} {\bibfield  {journal} {\bibinfo  {journal}
  {Philosophical Magazine}\ }\textbf {\bibinfo {volume} {25}},\ \bibinfo
  {pages} {1} (\bibinfo {year} {1972})}\BibitemShut {NoStop}%
\bibitem [{\citenamefont {Phillips}(1972)}]{PhillipsDistributionOfTLSs}%
  \BibitemOpen
  \bibfield  {author} {\bibinfo {author} {\bibfnamefont {W.}~\bibnamefont
  {Phillips}},\ }\href {\doibase 10.1007/BF00660072} {\bibfield  {journal}
  {\bibinfo  {journal} {Journal of Low Temperature Physics}\ }\textbf {\bibinfo
  {volume} {7}},\ \bibinfo {pages} {351} (\bibinfo {year} {1972})}\BibitemShut
  {NoStop}%
\bibitem [{\citenamefont {Phillips}(1987)}]{PhillipAmorphousLowTemp}%
  \BibitemOpen
  \bibfield  {author} {\bibinfo {author} {\bibfnamefont {W.~A.}\ \bibnamefont
  {Phillips}},\ }\href {http://stacks.iop.org/0034-4885/50/i=12/a=003}
  {\bibfield  {journal} {\bibinfo  {journal} {Reports on Progress in Physics}\
  }\textbf {\bibinfo {volume} {50}},\ \bibinfo {pages} {1657} (\bibinfo {year}
  {1987})}\BibitemShut {NoStop}%
\bibitem [{\citenamefont {Black}\ and\ \citenamefont
  {Halperin}(1977)}]{Black&HalperinTLS-TLSInt}%
  \BibitemOpen
  \bibfield  {author} {\bibinfo {author} {\bibfnamefont {J.~L.}\ \bibnamefont
  {Black}}\ and\ \bibinfo {author} {\bibfnamefont {B.~I.}\ \bibnamefont
  {Halperin}},\ }\href {\doibase 10.1103/PhysRevB.16.2879} {\bibfield
  {journal} {\bibinfo  {journal} {Phys. Rev. B}\ }\textbf {\bibinfo {volume}
  {16}},\ \bibinfo {pages} {2879} (\bibinfo {year} {1977})}\BibitemShut
  {NoStop}%
\bibitem [{\citenamefont {Burin}\ and\ \citenamefont
  {Kagan}(1994)}]{burinCrossOverTLS-TLSRelaxation}%
  \BibitemOpen
  \bibfield  {author} {\bibinfo {author} {\bibfnamefont {A.}~\bibnamefont
  {Burin}}\ and\ \bibinfo {author} {\bibfnamefont {Y.}~\bibnamefont {Kagan}},\
  }\href@noop {} {\bibfield  {journal} {\bibinfo  {journal} {Journal of
  Experimental and Theoretical Physics}\ } (\bibinfo {year}
  {1994})}\BibitemShut {NoStop}%
\bibitem [{\citenamefont {Rogge}\ \emph {et~al.}(1996)\citenamefont {Rogge},
  \citenamefont {Natelson},\ and\ \citenamefont
  {Osheroff}}]{Rogge&Natelson&Osheroff}%
  \BibitemOpen
  \bibfield  {author} {\bibinfo {author} {\bibfnamefont {S.}~\bibnamefont
  {Rogge}}, \bibinfo {author} {\bibfnamefont {D.}~\bibnamefont {Natelson}}, \
  and\ \bibinfo {author} {\bibfnamefont {D.~D.}\ \bibnamefont {Osheroff}},\
  }\href {\doibase 10.1103/PhysRevLett.76.3136} {\bibfield  {journal} {\bibinfo
   {journal} {Phys. Rev. Lett.}\ }\textbf {\bibinfo {volume} {76}},\ \bibinfo
  {pages} {3136} (\bibinfo {year} {1996})}\BibitemShut {NoStop}%
\bibitem [{\citenamefont {Natelson}\ \emph {et~al.}(1998)\citenamefont
  {Natelson}, \citenamefont {Rosenberg},\ and\ \citenamefont
  {Osheroff}}]{Natelson&Rosenberg&Osheroff}%
  \BibitemOpen
  \bibfield  {author} {\bibinfo {author} {\bibfnamefont {D.}~\bibnamefont
  {Natelson}}, \bibinfo {author} {\bibfnamefont {D.}~\bibnamefont {Rosenberg}},
  \ and\ \bibinfo {author} {\bibfnamefont {D.~D.}\ \bibnamefont {Osheroff}},\
  }\href {\doibase 10.1103/PhysRevLett.80.4689} {\bibfield  {journal} {\bibinfo
   {journal} {Phys. Rev. Lett.}\ }\textbf {\bibinfo {volume} {80}},\ \bibinfo
  {pages} {4689} (\bibinfo {year} {1998})}\BibitemShut {NoStop}%
\bibitem [{\citenamefont {Neeley}\ \emph {et~al.}(2008)\citenamefont {Neeley},
  \citenamefont {Ansmann}, \citenamefont {Bialczak}, \citenamefont {Hofheinz},
  \citenamefont {Katz}, \citenamefont {Lucero}, \citenamefont {O/'Connell},
  \citenamefont {Wang}, \citenamefont {Cleland},\ and\ \citenamefont
  {Martinis}}]{Neeley2008}%
  \BibitemOpen
  \bibfield  {author} {\bibinfo {author} {\bibfnamefont {M.}~\bibnamefont
  {Neeley}}, \bibinfo {author} {\bibfnamefont {M.}~\bibnamefont {Ansmann}},
  \bibinfo {author} {\bibfnamefont {R.~C.}\ \bibnamefont {Bialczak}}, \bibinfo
  {author} {\bibfnamefont {M.}~\bibnamefont {Hofheinz}}, \bibinfo {author}
  {\bibfnamefont {N.}~\bibnamefont {Katz}}, \bibinfo {author} {\bibfnamefont
  {E.}~\bibnamefont {Lucero}}, \bibinfo {author} {\bibfnamefont
  {A.}~\bibnamefont {O/'Connell}}, \bibinfo {author} {\bibfnamefont
  {H.}~\bibnamefont {Wang}}, \bibinfo {author} {\bibfnamefont {A.~N.}\
  \bibnamefont {Cleland}}, \ and\ \bibinfo {author} {\bibfnamefont {J.~M.}\
  \bibnamefont {Martinis}},\ }\href {\doibase 10.1038/nphys972} {\bibfield
  {journal} {\bibinfo  {journal} {Nat Phys}\ }\textbf {\bibinfo {volume} {4}},\
  \bibinfo {pages} {523} (\bibinfo {year} {2008})}\BibitemShut {NoStop}%
\bibitem [{\citenamefont {Simmonds}\ \emph {et~al.}(2004)\citenamefont
  {Simmonds}, \citenamefont {Lang}, \citenamefont {Hite}, \citenamefont {Nam},
  \citenamefont {Pappas},\ and\ \citenamefont {Martinis}}]{Simmonds2004}%
  \BibitemOpen
  \bibfield  {author} {\bibinfo {author} {\bibfnamefont {R.~W.}\ \bibnamefont
  {Simmonds}}, \bibinfo {author} {\bibfnamefont {K.~M.}\ \bibnamefont {Lang}},
  \bibinfo {author} {\bibfnamefont {D.~A.}\ \bibnamefont {Hite}}, \bibinfo
  {author} {\bibfnamefont {S.}~\bibnamefont {Nam}}, \bibinfo {author}
  {\bibfnamefont {D.~P.}\ \bibnamefont {Pappas}}, \ and\ \bibinfo {author}
  {\bibfnamefont {J.~M.}\ \bibnamefont {Martinis}},\ }\href {\doibase
  10.1103/PhysRevLett.93.077003} {\bibfield  {journal} {\bibinfo  {journal}
  {Phys. Rev. Lett.}\ }\textbf {\bibinfo {volume} {93}},\ \bibinfo {pages}
  {077003} (\bibinfo {year} {2004})}\BibitemShut {NoStop}%
\bibitem [{\citenamefont {Barends}\ \emph {et~al.}(2014)\citenamefont
  {Barends}, \citenamefont {Kelly}, \citenamefont {Megrant}, \citenamefont
  {Veitia}, \citenamefont {Sank}, \citenamefont {Jeffrey}, \citenamefont
  {White}, \citenamefont {Mutus}, \citenamefont {Fowler}, \citenamefont
  {Campbell}, \citenamefont {Chen}, \citenamefont {Chen}, \citenamefont
  {Chiaro}, \citenamefont {Dunsworth}, \citenamefont {Neill}, \citenamefont
  {O/'Malley}, \citenamefont {Roushan}, \citenamefont {Vainsencher},
  \citenamefont {Wenner}, \citenamefont {Korotkov}, \citenamefont {Cleland},\
  and\ \citenamefont {Martinis}}]{Barends2014}%
  \BibitemOpen
  \bibfield  {author} {\bibinfo {author} {\bibfnamefont {R.}~\bibnamefont
  {Barends}}, \bibinfo {author} {\bibfnamefont {J.}~\bibnamefont {Kelly}},
  \bibinfo {author} {\bibfnamefont {A.}~\bibnamefont {Megrant}}, \bibinfo
  {author} {\bibfnamefont {A.}~\bibnamefont {Veitia}}, \bibinfo {author}
  {\bibfnamefont {D.}~\bibnamefont {Sank}}, \bibinfo {author} {\bibfnamefont
  {E.}~\bibnamefont {Jeffrey}}, \bibinfo {author} {\bibfnamefont {T.~C.}\
  \bibnamefont {White}}, \bibinfo {author} {\bibfnamefont {J.}~\bibnamefont
  {Mutus}}, \bibinfo {author} {\bibfnamefont {A.~G.}\ \bibnamefont {Fowler}},
  \bibinfo {author} {\bibfnamefont {B.}~\bibnamefont {Campbell}}, \bibinfo
  {author} {\bibfnamefont {Y.}~\bibnamefont {Chen}}, \bibinfo {author}
  {\bibfnamefont {Z.}~\bibnamefont {Chen}}, \bibinfo {author} {\bibfnamefont
  {B.}~\bibnamefont {Chiaro}}, \bibinfo {author} {\bibfnamefont
  {A.}~\bibnamefont {Dunsworth}}, \bibinfo {author} {\bibfnamefont
  {C.}~\bibnamefont {Neill}}, \bibinfo {author} {\bibfnamefont
  {P.}~\bibnamefont {O/'Malley}}, \bibinfo {author} {\bibfnamefont
  {P.}~\bibnamefont {Roushan}}, \bibinfo {author} {\bibfnamefont
  {A.}~\bibnamefont {Vainsencher}}, \bibinfo {author} {\bibfnamefont
  {J.}~\bibnamefont {Wenner}}, \bibinfo {author} {\bibfnamefont {A.~N.}\
  \bibnamefont {Korotkov}}, \bibinfo {author} {\bibfnamefont {A.~N.}\
  \bibnamefont {Cleland}}, \ and\ \bibinfo {author} {\bibfnamefont {J.~M.}\
  \bibnamefont {Martinis}},\ }\href@noop {} {\bibfield  {journal} {\bibinfo
  {journal} {Nature}\ }\textbf {\bibinfo {volume} {508}},\ \bibinfo {pages}
  {500} (\bibinfo {year} {2014})},\ \bibinfo {note} {letter}\BibitemShut
  {NoStop}%
\bibitem [{\citenamefont {Shalibo}\ \emph {et~al.}(2010)\citenamefont
  {Shalibo}, \citenamefont {Rofe}, \citenamefont {Shwa}, \citenamefont
  {Zeides}, \citenamefont {Neeley}, \citenamefont {Martinis},\ and\
  \citenamefont {Katz}}]{Shalibo2010}%
  \BibitemOpen
  \bibfield  {author} {\bibinfo {author} {\bibfnamefont {Y.}~\bibnamefont
  {Shalibo}}, \bibinfo {author} {\bibfnamefont {Y.}~\bibnamefont {Rofe}},
  \bibinfo {author} {\bibfnamefont {D.}~\bibnamefont {Shwa}}, \bibinfo {author}
  {\bibfnamefont {F.}~\bibnamefont {Zeides}}, \bibinfo {author} {\bibfnamefont
  {M.}~\bibnamefont {Neeley}}, \bibinfo {author} {\bibfnamefont {J.~M.}\
  \bibnamefont {Martinis}}, \ and\ \bibinfo {author} {\bibfnamefont
  {N.}~\bibnamefont {Katz}},\ }\href {\doibase 10.1103/PhysRevLett.105.177001}
  {\bibfield  {journal} {\bibinfo  {journal} {Phys. Rev. Lett.}\ }\textbf
  {\bibinfo {volume} {105}},\ \bibinfo {pages} {177001} (\bibinfo {year}
  {2010})}\BibitemShut {NoStop}%
\bibitem [{\citenamefont {Lisenfeld}\ \emph {et~al.}(2016)\citenamefont
  {Lisenfeld}, \citenamefont {Bilmes}, \citenamefont {Matityahu}, \citenamefont
  {Zanker}, \citenamefont {Marthaler}, \citenamefont {Schechter}, \citenamefont
  {Sch{\"o}n}, \citenamefont {Shnirman}, \citenamefont {Weiss},\ and\
  \citenamefont {Ustinov}}]{lisenfeldDecoherenceTLSs}%
  \BibitemOpen
  \bibfield  {author} {\bibinfo {author} {\bibfnamefont {J.}~\bibnamefont
  {Lisenfeld}}, \bibinfo {author} {\bibfnamefont {A.}~\bibnamefont {Bilmes}},
  \bibinfo {author} {\bibfnamefont {S.}~\bibnamefont {Matityahu}}, \bibinfo
  {author} {\bibfnamefont {S.}~\bibnamefont {Zanker}}, \bibinfo {author}
  {\bibfnamefont {M.}~\bibnamefont {Marthaler}}, \bibinfo {author}
  {\bibfnamefont {M.}~\bibnamefont {Schechter}}, \bibinfo {author}
  {\bibfnamefont {G.}~\bibnamefont {Sch{\"o}n}}, \bibinfo {author}
  {\bibfnamefont {A.}~\bibnamefont {Shnirman}}, \bibinfo {author}
  {\bibfnamefont {G.}~\bibnamefont {Weiss}}, \ and\ \bibinfo {author}
  {\bibfnamefont {A.~V.}\ \bibnamefont {Ustinov}},\ }\href@noop {} {\bibfield
  {journal} {\bibinfo  {journal} {Scientific reports}\ }\textbf {\bibinfo
  {volume} {6}} (\bibinfo {year} {2016})}\BibitemShut {NoStop}%
\bibitem [{\citenamefont {Matityahu}\ \emph {et~al.}(2016)\citenamefont
  {Matityahu}, \citenamefont {Shnirman}, \citenamefont {Sch\"on},\ and\
  \citenamefont {Schechter}}]{Matityahu&Shnirman&Schechter}%
  \BibitemOpen
  \bibfield  {author} {\bibinfo {author} {\bibfnamefont {S.}~\bibnamefont
  {Matityahu}}, \bibinfo {author} {\bibfnamefont {A.}~\bibnamefont {Shnirman}},
  \bibinfo {author} {\bibfnamefont {G.}~\bibnamefont {Sch\"on}}, \ and\
  \bibinfo {author} {\bibfnamefont {M.}~\bibnamefont {Schechter}},\ }\href
  {\doibase 10.1103/PhysRevB.93.134208} {\bibfield  {journal} {\bibinfo
  {journal} {Phys. Rev. B}\ }\textbf {\bibinfo {volume} {93}},\ \bibinfo
  {pages} {134208} (\bibinfo {year} {2016})}\BibitemShut {NoStop}%
\bibitem [{\citenamefont {Lisenfeld}\ \emph {et~al.}(2015)\citenamefont
  {Lisenfeld}, \citenamefont {Grabovskij}, \citenamefont {M{\"u}ller},
  \citenamefont {Cole}, \citenamefont {Weiss},\ and\ \citenamefont
  {Ustinov}}]{Lisenfeld2015}%
  \BibitemOpen
  \bibfield  {author} {\bibinfo {author} {\bibfnamefont {J.}~\bibnamefont
  {Lisenfeld}}, \bibinfo {author} {\bibfnamefont {G.~J.}\ \bibnamefont
  {Grabovskij}}, \bibinfo {author} {\bibfnamefont {C.}~\bibnamefont
  {M{\"u}ller}}, \bibinfo {author} {\bibfnamefont {J.~H.}\ \bibnamefont
  {Cole}}, \bibinfo {author} {\bibfnamefont {G.}~\bibnamefont {Weiss}}, \ and\
  \bibinfo {author} {\bibfnamefont {A.~V.}\ \bibnamefont {Ustinov}},\
  }\href@noop {} {\bibfield  {journal} {\bibinfo  {journal} {Nat Commun}\
  }\textbf {\bibinfo {volume} {6}} (\bibinfo {year} {2015})}\BibitemShut
  {NoStop}%
\bibitem [{\citenamefont {Lisenfeld}\ \emph {et~al.}(2010)\citenamefont
  {Lisenfeld}, \citenamefont {M\"uller}, \citenamefont {Cole}, \citenamefont
  {Bushev}, \citenamefont {Lukashenko}, \citenamefont {Shnirman},\ and\
  \citenamefont {Ustinov}}]{LisenfeldTDependenceOfSingleTLS}%
  \BibitemOpen
  \bibfield  {author} {\bibinfo {author} {\bibfnamefont {J.}~\bibnamefont
  {Lisenfeld}}, \bibinfo {author} {\bibfnamefont {C.}~\bibnamefont {M\"uller}},
  \bibinfo {author} {\bibfnamefont {J.~H.}\ \bibnamefont {Cole}}, \bibinfo
  {author} {\bibfnamefont {P.}~\bibnamefont {Bushev}}, \bibinfo {author}
  {\bibfnamefont {A.}~\bibnamefont {Lukashenko}}, \bibinfo {author}
  {\bibfnamefont {A.}~\bibnamefont {Shnirman}}, \ and\ \bibinfo {author}
  {\bibfnamefont {A.~V.}\ \bibnamefont {Ustinov}},\ }\href {\doibase
  10.1103/PhysRevLett.105.230504} {\bibfield  {journal} {\bibinfo  {journal}
  {Phys. Rev. Lett.}\ }\textbf {\bibinfo {volume} {105}},\ \bibinfo {pages}
  {230504} (\bibinfo {year} {2010})}\BibitemShut {NoStop}%
\bibitem [{\citenamefont {Grabovskij}\ \emph {et~al.}(2012)\citenamefont
  {Grabovskij}, \citenamefont {Peichl}, \citenamefont {Lisenfeld},
  \citenamefont {Weiss},\ and\ \citenamefont
  {Ustinov}}]{Grabovskij&LisenfeldStrainTuningOfSingleTLS}%
  \BibitemOpen
  \bibfield  {author} {\bibinfo {author} {\bibfnamefont {G.~J.}\ \bibnamefont
  {Grabovskij}}, \bibinfo {author} {\bibfnamefont {T.}~\bibnamefont {Peichl}},
  \bibinfo {author} {\bibfnamefont {J.}~\bibnamefont {Lisenfeld}}, \bibinfo
  {author} {\bibfnamefont {G.}~\bibnamefont {Weiss}}, \ and\ \bibinfo {author}
  {\bibfnamefont {A.~V.}\ \bibnamefont {Ustinov}},\ }\href {\doibase
  10.1126/science.1226487} {\bibfield  {journal} {\bibinfo  {journal}
  {Science}\ }\textbf {\bibinfo {volume} {338}},\ \bibinfo {pages} {232}
  (\bibinfo {year} {2012})}\BibitemShut {NoStop}%
\bibitem [{\citenamefont {M\"uller}\ \emph {et~al.}(2015)\citenamefont
  {M\"uller}, \citenamefont {Lisenfeld}, \citenamefont {Shnirman},\ and\
  \citenamefont {Poletto}}]{Muller&LisenfeldNoiseFromSingleTLS}%
  \BibitemOpen
  \bibfield  {author} {\bibinfo {author} {\bibfnamefont {C.}~\bibnamefont
  {M\"uller}}, \bibinfo {author} {\bibfnamefont {J.}~\bibnamefont {Lisenfeld}},
  \bibinfo {author} {\bibfnamefont {A.}~\bibnamefont {Shnirman}}, \ and\
  \bibinfo {author} {\bibfnamefont {S.}~\bibnamefont {Poletto}},\ }\href
  {\doibase 10.1103/PhysRevB.92.035442} {\bibfield  {journal} {\bibinfo
  {journal} {Phys. Rev. B}\ }\textbf {\bibinfo {volume} {92}},\ \bibinfo
  {pages} {035442} (\bibinfo {year} {2015})}\BibitemShut {NoStop}%
\bibitem [{\citenamefont {Burin}(1995)}]{Burin'sDipoleGap}%
  \BibitemOpen
  \bibfield  {author} {\bibinfo {author} {\bibfnamefont {A.}~\bibnamefont
  {Burin}},\ }\href {\doibase 10.1007/BF00751512} {\bibfield  {journal}
  {\bibinfo  {journal} {Journal of Low Temperature Physics}\ }\textbf {\bibinfo
  {volume} {100}},\ \bibinfo {pages} {309} (\bibinfo {year}
  {1995})}\BibitemShut {NoStop}%
\bibitem [{\citenamefont {Burin}\ \emph {et~al.}(1998)\citenamefont {Burin},
  \citenamefont {Natelson}, \citenamefont {Osheroff},\ and\ \citenamefont
  {Kagan}}]{Burin1998}%
  \BibitemOpen
  \bibfield  {author} {\bibinfo {author} {\bibfnamefont {A.~L.}\ \bibnamefont
  {Burin}}, \bibinfo {author} {\bibfnamefont {D.}~\bibnamefont {Natelson}},
  \bibinfo {author} {\bibfnamefont {D.~D.}\ \bibnamefont {Osheroff}}, \ and\
  \bibinfo {author} {\bibfnamefont {Y.}~\bibnamefont {Kagan}},\ }in\ \href@noop
  {} {\emph {\bibinfo {booktitle} {Tunneling Systems in Amorphous and
  Crystalline Solids}}}\ (\bibinfo  {publisher} {Springer},\ \bibinfo {year}
  {1998})\ pp.\ \bibinfo {pages} {223--315}\BibitemShut {NoStop}%
\bibitem [{\citenamefont {Burin}\ and\ \citenamefont
  {Kagan}(1996)}]{Burin&Kagan}%
  \BibitemOpen
  \bibfield  {author} {\bibinfo {author} {\bibfnamefont {A.}~\bibnamefont
  {Burin}}\ and\ \bibinfo {author} {\bibfnamefont {Y.}~\bibnamefont {Kagan}},\
  }\href@noop {} {\bibfield  {journal} {\bibinfo  {journal} {Physics Letters
  A}\ }\textbf {\bibinfo {volume} {215}},\ \bibinfo {pages} {191 } (\bibinfo
  {year} {1996})}\BibitemShut {NoStop}%
\bibitem [{\citenamefont {Amir}\ \emph {et~al.}(2008)\citenamefont {Amir},
  \citenamefont {Oreg},\ and\ \citenamefont {Imry}}]{Ariel'sPaper}%
  \BibitemOpen
  \bibfield  {author} {\bibinfo {author} {\bibfnamefont {A.}~\bibnamefont
  {Amir}}, \bibinfo {author} {\bibfnamefont {Y.}~\bibnamefont {Oreg}}, \ and\
  \bibinfo {author} {\bibfnamefont {Y.}~\bibnamefont {Imry}},\ }\href {\doibase
  10.1103/PhysRevB.77.165207} {\bibfield  {journal} {\bibinfo  {journal} {Phys.
  Rev. B}\ }\textbf {\bibinfo {volume} {77}},\ \bibinfo {pages} {165207}
  (\bibinfo {year} {2008})}\BibitemShut {NoStop}%
\bibitem [{\citenamefont {Churkin}\ \emph {et~al.}(2014)\citenamefont
  {Churkin}, \citenamefont {Barash},\ and\ \citenamefont
  {Schechter}}]{ChurkinTLSDOSMonteCarlo}%
  \BibitemOpen
  \bibfield  {author} {\bibinfo {author} {\bibfnamefont {A.}~\bibnamefont
  {Churkin}}, \bibinfo {author} {\bibfnamefont {D.}~\bibnamefont {Barash}}, \
  and\ \bibinfo {author} {\bibfnamefont {M.}~\bibnamefont {Schechter}},\ }\href
  {\doibase 10.1103/PhysRevB.89.104202} {\bibfield  {journal} {\bibinfo
  {journal} {Phys. Rev. B}\ }\textbf {\bibinfo {volume} {89}},\ \bibinfo
  {pages} {104202} (\bibinfo {year} {2014})}\BibitemShut {NoStop}%
\bibitem [{\citenamefont {Schechter}\ and\ \citenamefont
  {Stamp}(2013)}]{MosheStrong&WeakTLSs}%
  \BibitemOpen
  \bibfield  {author} {\bibinfo {author} {\bibfnamefont {M.}~\bibnamefont
  {Schechter}}\ and\ \bibinfo {author} {\bibfnamefont {P.~C.~E.}\ \bibnamefont
  {Stamp}},\ }\href {\doibase 10.1103/PhysRevB.88.174202} {\bibfield  {journal}
  {\bibinfo  {journal} {Phys. Rev. B}\ }\textbf {\bibinfo {volume} {88}},\
  \bibinfo {pages} {174202} (\bibinfo {year} {2013})}\BibitemShut {NoStop}%
\bibitem [{\citenamefont {Churkin}\ \emph {et~al.}(2013)\citenamefont
  {Churkin}, \citenamefont {Gabdank}, \citenamefont {Burin},\ and\
  \citenamefont {Schechter}}]{churkin2013strain}%
  \BibitemOpen
  \bibfield  {author} {\bibinfo {author} {\bibfnamefont {A.}~\bibnamefont
  {Churkin}}, \bibinfo {author} {\bibfnamefont {I.}~\bibnamefont {Gabdank}},
  \bibinfo {author} {\bibfnamefont {A.}~\bibnamefont {Burin}}, \ and\ \bibinfo
  {author} {\bibfnamefont {M.}~\bibnamefont {Schechter}},\ }\href@noop {}
  {\bibfield  {journal} {\bibinfo  {journal} {arXiv preprint arXiv:1307.0868}\
  } (\bibinfo {year} {2013})}\BibitemShut {NoStop}%
\bibitem [{\citenamefont {Baranovskii}\ \emph {et~al.}(1980)\citenamefont
  {Baranovskii}, \citenamefont {Shklovskii},\ and\ \citenamefont
  {Efros}}]{CoulombAndDipoleGapsDerivation}%
  \BibitemOpen
  \bibfield  {author} {\bibinfo {author} {\bibfnamefont {S.}~\bibnamefont
  {Baranovskii}}, \bibinfo {author} {\bibfnamefont {B.}~\bibnamefont
  {Shklovskii}}, \ and\ \bibinfo {author} {\bibfnamefont {A.}~\bibnamefont
  {Efros}},\ }\href@noop {} {\bibfield  {journal} {\bibinfo  {journal} {Soviet
  Journal of Experimental and Theoretical Physics}\ }\textbf {\bibinfo {volume}
  {51}},\ \bibinfo {pages} {199} (\bibinfo {year} {1980})}\BibitemShut
  {NoStop}%
\bibitem [{\citenamefont {Cuevas}\ \emph {et~al.}(1989)\citenamefont {Cuevas},
  \citenamefont {Chic{\'o}n},\ and\ \citenamefont
  {Ortu{\~n}o}}]{CuevasTLSDOSMonteCarlo}%
  \BibitemOpen
  \bibfield  {author} {\bibinfo {author} {\bibfnamefont {E.}~\bibnamefont
  {Cuevas}}, \bibinfo {author} {\bibfnamefont {R.}~\bibnamefont {Chic{\'o}n}},
  \ and\ \bibinfo {author} {\bibfnamefont {M.}~\bibnamefont {Ortu{\~n}o}},\
  }\href@noop {} {\bibfield  {journal} {\bibinfo  {journal} {Physica B:
  Condensed Matter}\ }\textbf {\bibinfo {volume} {160}},\ \bibinfo {pages}
  {293} (\bibinfo {year} {1989})}\BibitemShut {NoStop}%
\bibitem [{\citenamefont {Efros}\ and\ \citenamefont
  {Shklovskii}(1975)}]{EfrosShklovskiiCoulombGapT=0}%
  \BibitemOpen
  \bibfield  {author} {\bibinfo {author} {\bibfnamefont {A.~L.}\ \bibnamefont
  {Efros}}\ and\ \bibinfo {author} {\bibfnamefont {B.~I.}\ \bibnamefont
  {Shklovskii}},\ }\href@noop {} {\bibfield  {journal} {\bibinfo  {journal}
  {Journal of Physics C: Solid State Physics}\ }\textbf {\bibinfo {volume}
  {8}},\ \bibinfo {pages} {L49} (\bibinfo {year} {1975})}\BibitemShut {NoStop}%
\bibitem [{\citenamefont {Grunewald}\ \emph {et~al.}(1982)\citenamefont
  {Grunewald}, \citenamefont {Pohlmann}, \citenamefont {Schweitzer},\ and\
  \citenamefont {Wurtz}}]{NumericalMethodForDOS}%
  \BibitemOpen
  \bibfield  {author} {\bibinfo {author} {\bibfnamefont {M.}~\bibnamefont
  {Grunewald}}, \bibinfo {author} {\bibfnamefont {B.}~\bibnamefont {Pohlmann}},
  \bibinfo {author} {\bibfnamefont {L.}~\bibnamefont {Schweitzer}}, \ and\
  \bibinfo {author} {\bibfnamefont {D.}~\bibnamefont {Wurtz}},\ }\href@noop {}
  {\bibfield  {journal} {\bibinfo  {journal} {Journal of Physics C: Solid State
  Physics}\ }\textbf {\bibinfo {volume} {15}},\ \bibinfo {pages} {L1153}
  (\bibinfo {year} {1982})}\BibitemShut {NoStop}%
\bibitem [{\citenamefont {J{\"a}ckle}(1972)}]{JakleTransitionRatesOfTLSs}%
  \BibitemOpen
  \bibfield  {author} {\bibinfo {author} {\bibfnamefont {J.}~\bibnamefont
  {J{\"a}ckle}},\ }\href {\doibase 10.1007/BF01401204} {\bibfield  {journal}
  {\bibinfo  {journal} {Zeitschrift f{\"u}r Physik}\ }\textbf {\bibinfo
  {volume} {257}},\ \bibinfo {pages} {212} (\bibinfo {year}
  {1972})}\BibitemShut {NoStop}%
\bibitem [{\citenamefont {Bender}\ and\ \citenamefont {Orszag}(1999)}]{ExpInt}%
  \BibitemOpen
  \bibfield  {author} {\bibinfo {author} {\bibfnamefont {C.~M.}\ \bibnamefont
  {Bender}}\ and\ \bibinfo {author} {\bibfnamefont {S.~A.}\ \bibnamefont
  {Orszag}},\ }\href@noop {} {\emph {\bibinfo {title} {{Advanced Mathematical
  Methods for Scientists and Engineers: Asymptotic methods and perturbation
  theory}}}}\ (\bibinfo  {publisher} {Springer},\ \bibinfo {year} {1999})\ p.\
  \bibinfo {pages} {252}\BibitemShut {NoStop}%
\bibitem [{\citenamefont {Amir}\ \emph {et~al.}(2012)\citenamefont {Amir},
  \citenamefont {Oreg},\ and\ \citenamefont {Imry}}]{AmirPnas2012}%
  \BibitemOpen
  \bibfield  {author} {\bibinfo {author} {\bibfnamefont {A.}~\bibnamefont
  {Amir}}, \bibinfo {author} {\bibfnamefont {Y.}~\bibnamefont {Oreg}}, \ and\
  \bibinfo {author} {\bibfnamefont {Y.}~\bibnamefont {Imry}},\ }\href {\doibase
  10.1073/pnas.1120147109} {\bibfield  {journal} {\bibinfo  {journal}
  {Proceedings of the National Academy of Sciences}\ }\textbf {\bibinfo
  {volume} {109}},\ \bibinfo {pages} {1850} (\bibinfo {year}
  {2012})}\BibitemShut {NoStop}%
\bibitem [{\citenamefont {Burin}\ \emph {et~al.}(2006)\citenamefont {Burin},
  \citenamefont {Shklovskii}, \citenamefont {Kozub}, \citenamefont {Galperin},\
  and\ \citenamefont {Vinokur}}]{BurinElectronGlass}%
  \BibitemOpen
  \bibfield  {author} {\bibinfo {author} {\bibfnamefont {A.~L.}\ \bibnamefont
  {Burin}}, \bibinfo {author} {\bibfnamefont {B.~I.}\ \bibnamefont
  {Shklovskii}}, \bibinfo {author} {\bibfnamefont {V.~I.}\ \bibnamefont
  {Kozub}}, \bibinfo {author} {\bibfnamefont {Y.~M.}\ \bibnamefont {Galperin}},
  \ and\ \bibinfo {author} {\bibfnamefont {V.}~\bibnamefont {Vinokur}},\ }\href
  {\doibase 10.1103/PhysRevB.74.075205} {\bibfield  {journal} {\bibinfo
  {journal} {Phys. Rev. B}\ }\textbf {\bibinfo {volume} {74}},\ \bibinfo
  {pages} {075205} (\bibinfo {year} {2006})}\BibitemShut {NoStop}%
\bibitem [{\citenamefont {Amir}\ \emph {et~al.}(2011)\citenamefont {Amir},
  \citenamefont {Oreg},\ and\ \citenamefont {Imry}}]{2011ArielReview}%
  \BibitemOpen
  \bibfield  {author} {\bibinfo {author} {\bibfnamefont {A.}~\bibnamefont
  {Amir}}, \bibinfo {author} {\bibfnamefont {Y.}~\bibnamefont {Oreg}}, \ and\
  \bibinfo {author} {\bibfnamefont {Y.}~\bibnamefont {Imry}},\ }\href {\doibase
  10.1146/annurev-conmatphys-062910-140455} {\bibfield  {journal} {\bibinfo
  {journal} {Annual Review of Condensed Matter Physics}\ }\textbf {\bibinfo
  {volume} {2}},\ \bibinfo {pages} {235} (\bibinfo {year} {2011})}\BibitemShut
  {NoStop}%
\bibitem [{\citenamefont {Pollak}\ \emph {et~al.}(2012)\citenamefont {Pollak},
  \citenamefont {Ortu{\~ n}o},\ and\ \citenamefont {Frydman}}]{pollak2012}%
  \BibitemOpen
  \bibfield  {author} {\bibinfo {author} {\bibfnamefont {M.}~\bibnamefont
  {Pollak}}, \bibinfo {author} {\bibfnamefont {M.}~\bibnamefont {Ortu{\~ n}o}},
  \ and\ \bibinfo {author} {\bibfnamefont {A.}~\bibnamefont {Frydman}},\ }\href
  {\doibase 10.1017/CBO9780511978999} {\emph {\bibinfo {title} {The Electron
  Glass:}}}\ (\bibinfo  {publisher} {Cambridge University Press},\ \bibinfo
  {address} {Cambridge},\ \bibinfo {year} {2012})\BibitemShut {NoStop}%
\bibitem [{\citenamefont {Davies}\ \emph {et~al.}(1984)\citenamefont {Davies},
  \citenamefont {Lee},\ and\ \citenamefont {Rice}}]{DaviesCoulombGapT>0}%
  \BibitemOpen
  \bibfield  {author} {\bibinfo {author} {\bibfnamefont {J.~H.}\ \bibnamefont
  {Davies}}, \bibinfo {author} {\bibfnamefont {P.~A.}\ \bibnamefont {Lee}}, \
  and\ \bibinfo {author} {\bibfnamefont {T.~M.}\ \bibnamefont {Rice}},\ }\href
  {\doibase 10.1103/PhysRevB.29.4260} {\bibfield  {journal} {\bibinfo
  {journal} {Phys. Rev. B}\ }\textbf {\bibinfo {volume} {29}},\ \bibinfo
  {pages} {4260} (\bibinfo {year} {1984})}\BibitemShut {NoStop}%
\bibitem [{\citenamefont {Levin}\ \emph {et~al.}(1987)\citenamefont {Levin},
  \citenamefont {Nguen}, \citenamefont {Shklovsii},\ and\ \citenamefont
  {Efros}}]{LevinCoulombGapT>0}%
  \BibitemOpen
  \bibfield  {author} {\bibinfo {author} {\bibfnamefont {E.}~\bibnamefont
  {Levin}}, \bibinfo {author} {\bibfnamefont {V.}~\bibnamefont {Nguen}},
  \bibinfo {author} {\bibfnamefont {B.}~\bibnamefont {Shklovsii}}, \ and\
  \bibinfo {author} {\bibfnamefont {A.}~\bibnamefont {Efros}},\ }\href@noop {}
  {\bibfield  {journal} {\bibinfo  {journal} {Soviet Physics JETP}\ }\textbf
  {\bibinfo {volume} {65}},\ \bibinfo {pages} {842} (\bibinfo {year}
  {1987})}\BibitemShut {NoStop}%
\bibitem [{\citenamefont {Pikus}\ and\ \citenamefont
  {Efros}(1994)}]{PikusCoulombGapT>0}%
  \BibitemOpen
  \bibfield  {author} {\bibinfo {author} {\bibfnamefont {F.~G.}\ \bibnamefont
  {Pikus}}\ and\ \bibinfo {author} {\bibfnamefont {A.~L.}\ \bibnamefont
  {Efros}},\ }\href {\doibase 10.1103/PhysRevLett.73.3014} {\bibfield
  {journal} {\bibinfo  {journal} {Phys. Rev. Lett.}\ }\textbf {\bibinfo
  {volume} {73}},\ \bibinfo {pages} {3014} (\bibinfo {year}
  {1994})}\BibitemShut {NoStop}%
\bibitem [{\citenamefont {Miller}\ and\ \citenamefont
  {Abrahams}(1960)}]{Miller-Abrahams-Transitions}%
  \BibitemOpen
  \bibfield  {author} {\bibinfo {author} {\bibfnamefont {A.}~\bibnamefont
  {Miller}}\ and\ \bibinfo {author} {\bibfnamefont {E.}~\bibnamefont
  {Abrahams}},\ }\href {\doibase 10.1103/PhysRev.120.745} {\bibfield  {journal}
  {\bibinfo  {journal} {Phys. Rev.}\ }\textbf {\bibinfo {volume} {120}},\
  \bibinfo {pages} {745} (\bibinfo {year} {1960})}\BibitemShut {NoStop}%
\bibitem [{\citenamefont {Amir}\ \emph
  {et~al.}(2009{\natexlab{a}})\citenamefont {Amir}, \citenamefont {Oreg},\ and\
  \citenamefont {Imry}}]{PhysRevB.80.245214}%
  \BibitemOpen
  \bibfield  {author} {\bibinfo {author} {\bibfnamefont {A.}~\bibnamefont
  {Amir}}, \bibinfo {author} {\bibfnamefont {Y.}~\bibnamefont {Oreg}}, \ and\
  \bibinfo {author} {\bibfnamefont {Y.}~\bibnamefont {Imry}},\ }\href {\doibase
  10.1103/PhysRevB.80.245214} {\bibfield  {journal} {\bibinfo  {journal} {Phys.
  Rev. B}\ }\textbf {\bibinfo {volume} {80}},\ \bibinfo {pages} {245214}
  (\bibinfo {year} {2009}{\natexlab{a}})}\BibitemShut {NoStop}%
\bibitem [{\citenamefont {Amir}\ \emph {et~al.}(2010)\citenamefont {Amir},
  \citenamefont {Oreg},\ and\ \citenamefont {Imry}}]{PhysRevLett.105.070601}%
  \BibitemOpen
  \bibfield  {author} {\bibinfo {author} {\bibfnamefont {A.}~\bibnamefont
  {Amir}}, \bibinfo {author} {\bibfnamefont {Y.}~\bibnamefont {Oreg}}, \ and\
  \bibinfo {author} {\bibfnamefont {Y.}~\bibnamefont {Imry}},\ }\href {\doibase
  10.1103/PhysRevLett.105.070601} {\bibfield  {journal} {\bibinfo  {journal}
  {Phys. Rev. Lett.}\ }\textbf {\bibinfo {volume} {105}},\ \bibinfo {pages}
  {070601} (\bibinfo {year} {2010})}\BibitemShut {NoStop}%
\bibitem [{\citenamefont {Vaknin}\ \emph {et~al.}(2000)\citenamefont {Vaknin},
  \citenamefont {Ovadyahu},\ and\ \citenamefont {Pollak}}]{AgingEffectsVaknin}%
  \BibitemOpen
  \bibfield  {author} {\bibinfo {author} {\bibfnamefont {A.}~\bibnamefont
  {Vaknin}}, \bibinfo {author} {\bibfnamefont {Z.}~\bibnamefont {Ovadyahu}}, \
  and\ \bibinfo {author} {\bibfnamefont {M.}~\bibnamefont {Pollak}},\ }\href
  {\doibase 10.1103/PhysRevLett.84.3402} {\bibfield  {journal} {\bibinfo
  {journal} {Phys. Rev. Lett.}\ }\textbf {\bibinfo {volume} {84}},\ \bibinfo
  {pages} {3402} (\bibinfo {year} {2000})}\BibitemShut {NoStop}%
\bibitem [{\citenamefont {Amir}\ \emph
  {et~al.}(2009{\natexlab{b}})\citenamefont {Amir}, \citenamefont {Oreg},\ and\
  \citenamefont {Imry}}]{SlowRelaxationsAmir}%
  \BibitemOpen
  \bibfield  {author} {\bibinfo {author} {\bibfnamefont {A.}~\bibnamefont
  {Amir}}, \bibinfo {author} {\bibfnamefont {Y.}~\bibnamefont {Oreg}}, \ and\
  \bibinfo {author} {\bibfnamefont {Y.}~\bibnamefont {Imry}},\ }\href {\doibase
  10.1103/PhysRevLett.103.126403} {\bibfield  {journal} {\bibinfo  {journal}
  {Phys. Rev. Lett.}\ }\textbf {\bibinfo {volume} {103}},\ \bibinfo {pages}
  {126403} (\bibinfo {year} {2009}{\natexlab{b}})}\BibitemShut {NoStop}%
\end{thebibliography}%

\end{document}